\documentclass[prb,twocolumn,floatfix,amsmath]{revtex4-1}

\usepackage{graphicx}
\usepackage{color}
\usepackage{dsfont}
\usepackage[normalem]{ulem}
\usepackage{dcolumn}

\newcommand{\vect}[1]{\ensuremath{\mathbf{#1}}}
\newcommand{\matr}[1]{\ensuremath{\mathcal{#1}}}
\newcommand{\T}{\ensuremath{\mathrm{T}}}

\newcommand{\dx}{\ensuremath{\mathrm{d}x}}
\newcommand{\dd}{\ensuremath{\mathrm{d}}}

\newcommand{\uvect}[1]{\ensuremath{\hat{\mathbf{#1}}}}
\newcommand{\expo}[1]{\ensuremath{\mathrm{e}^{#1}}}
\newcommand{\abs}[1]{\ensuremath{\lvert #1 \rvert}}
\newcommand{\ket}[1]{\ensuremath{\lvert #1 \rangle}}

\definecolor{orange}{rgb}{1,0.5,0}
\definecolor{darkgreen}{RGB}{0,100,0}

\DeclareMathOperator{\sign}{sign}
\DeclareMathOperator{\am}{am}

\newcommand{\etal}{\textit{et al.\@ }}

\newcolumntype{d}[1]{D{.}{.}{#1}}

\makeatother

\begin{document}

\title{First-principles analysis of a homo-chiral cycloidal magnetic structure in a monolayer Cr on W(110)}
\author{Bernd Zimmermann}\email[Corresponding author: ]{be.zimmermann@fz-juelich.de} 
\author{Marcus Heide}
\author{Gustav Bihlmayer}
\author{Stefan Bl\"ugel}
\affiliation{Peter Gr\"unberg Institut and Institute for Advanced Simulation, Forschungszentrum J\"ulich and JARA, 52425 J\"ulich, Germany}

\date{\today}

\begin{abstract}
The magnetic structure of a Cr monolayer on a W(110) substrate is investigated by means of first-principles calculations based on the noncollinear spin density functional theory (DFT). As magnetic ground state we find a long-period homochiral  left-rotating spin spiral on-top of an atomic-scale anti-ferromagnetic order of nearest neighbor atoms. The rotation angle of the magnetic moment changes inhomogeneously from atom to atom across the spiral. We predict a propagation direction along the crystallographic $[001]$ direction with a period length of $\abs{\lambda} = 14.3~\mathrm{nm}$, which is in excellent agreement with a modulation of the local anti-ferromagnetic contrast observed in spin-polarized scanning tunneling microscope experiments by Santos \etal [New~J.~Phys.\ \textbf{10}, 013005 (2008)].
We identify the Dzyaloshinskii-Moriya interaction (DMI) as origin of the homochiral magnetic structure, competing with the Heisenberg-type exchange interaction and magneto-crystalline anisotropy energy. From DFT calculations we extract parameters for a micromagnetic model and thereby determine a considerable inhomogeneity of the spin spiral, increasing the period length by 6\% compared to homogeneous spin spirals. The results are compared to the behavior of a Mn and Fe monolayer and Fe doublelayer on a W(110) substrate.
\end{abstract}

\maketitle

\section{Introduction}
The observation of a new magnetic phase in an atomic monolayer of Mn on a W(110) substrate,\cite{Bode:07.1} whose magnetic ground state is a frozen cycloid of unique rotational sense, has opened a completely new vista of thin film magnetism. The occurrence of such a homochiral magnetic structure gives evidence of a sizeable antisymmetric exchange interaction, known as Dzyaloshinskii-Moriya interaction (DMI),\cite{Dzyaloshinskii:57.1, Moriya:60.1} an interaction that was so far basically ignored in the field of low-dimensional metallic magnets.\cite{BlandHeinrich, Bluegel:07.1} The DMI is a relativistic effect and spin-orbit coupling (SOC) is crucial for its occurrence. It arises due to the propagation of electrons in an inversion asymmetric environment. The observation of the chiral magnetic order in the Mn monolayer gives evidence that the DMI is of a size enabling a competition with other important magnetic interactions such as the Heisenberg exchange or the magneto-crystalline anisotropy energy and gives rise to spiraling magnetic ground-state structures. 

In fact, after the investigation of Mn on W(110), a few additional magnetic thin-film systems deposited on heavy element substrates exhibiting a chiral magnetic order have been investigated, {\it e.g.}\ Mn on W(100),\cite{Ferriani:08.1} a Mn doublelayer on W(110),\cite{Yosida:2012} Fe on Ir(111)\cite{Heinze:11} or a Pd-Fe doublelayer on Ir(111),\cite{Romming:13,Dupe:2014} but also past systems had been reinvestigated. 
One example is the analysis of domain-walls in stripes of Fe doublelayers on W(110), for which  Kubetzka \etal reported in Ref.~\onlinecite{Kubetzka:02.1} the surprising observation of  dense stripe domains with a defined sense of magnetic rotation. In the light of the Mn/W(110) experiments, Heide \etal\cite{Heide:08.1} explained these experiments in terms of a right-rotating chiral N\'eel-type domain wall, whose sense of rotation in the wall, the orientation of the wall relative to the lattice and the type of  wall was determined by the DMI. This theoretical analysis could be confirmed experimentally using spin-polarized scanning tunneling microscopy (SP-STM) performed in a triple axes vector magnet.\cite{Meckler:09.1} Moreover, a chiral asymmetry of the magnon dispersion due to the DMI was predicted by first-principles calculations for an Fe monolayer on W(110)\cite{Udvardi:2009} and measured in an Fe doublelayer on W(110).\cite{Zakeri:2010}

The observation of a large DMI and the formation of chiral magnetic structures are particularly influential for the field of spintronics: (i)  Chiral domain walls are much more stable against the Walker break-down and provide in conjunction with the current induced domain-wall motion a new opportunity for the realisation of the race-track memory.\cite{Thiaville:12,Ryu:13,Emori:13} (ii) The DMI is a basic ingredient for the formation of topological solitons in magnets, so-called magnetic skyrmions, that are currently being explored in thin film systems~\cite{Heinze:11,Romming:13,Dupe:2014} as a possible new magnetic particle for information technology.\cite{Fert:13} (iii) Electrons propagating along such winding magnetic structures accumulate Berry phases that translate into transport properties {\it e.g.}\ the topological Hall effect~\cite{Neubauer:09,Franz:14} arising from large emergent electrical and magnetic fields or give rise to additional spin torques~\cite{Schulz:12} enabling new design principles of magnetic devices.

In this paper we return to a Cr monolayer on W(110) that has been investigated by Santos {\it et al.}.\cite{Santos:08.1} Combining SP-STM experiments with {\it ab initio} calculations has shown that Cr exhibits a checkerboard type arrangement of magnetic moments coupling antiferromagnetically between nearest neighbor atoms. Further experiments revealed on top of this atomic-scale antiferromagnetic c(2$\times$2)-structure a long-period modulation along the [001]-direction with periodically repeated lines of blurred magnetic contrast every $7.7\pm 0.5$~nm. This magnetic structure was not further resolved, neither theoretically nor experimentally, but it reminds at similar findings observed for Mn/W(110)\cite{Bode:07.1} whose magnetic ground state was identified as a homo-chiral left-rotating cycloidal spin spiral along the [$1\bar{1}0$] direction with an experimental period length of 12~nm and thus vanishing SP-STM contrast every 6~nm. 
This difference in modulation direction is somehow at odds with the conventional working hypothesis accepted by a wide community that the DMI is determined by the element with strong spin-orbit interaction at the interface, and thus W(110) should be the key to the same modulation direction and the same rotational sense for both systems that are so similar.
On the other hand, in difference to Mn, Cr is also known to form a frozen sinusoidal spin-density wave (SDW) as bulk solid\cite{Schiller:04} and on the (110) surface.\cite{Santos:08.1}

The aim of this paper is to resolve the ground-state magnetic structure of a Cr monolayer on W(110) using a multi-scale approach. We first perform DFT total-energy calculations of non-collinear magnetic states that are described by flat homogeneous spin spirals. The calculations are carried out employing the generalized Bloch theorem,\cite{Sandratskii:91.1} which allows to calculate the magnetic structure for an arbitrary spin-spiral vector $\vect{q}$ on the basis of the chemical, {\it i.e.}\ p(1$\times$1) unit cell. This procedure is very time-saving but works only as long as the spin-orbit interaction is neglected. A value of homogeneous spin spirals lies in the observation that they are also solutions of the classical Heisenberg model for periodic lattices, which is typically the interaction with the largest energy scale in any spin-model. From the comparison of the total-energy calculations to the Heisenberg model one can conclude that for the case of Cr on W(110), the Heisenberg model catches all essential exchange-caused spin-interactions and leads to the conclusion that, in difference to Fe/Ir(111) \cite{Heinze:11} or a Mn doublelayer on W(110),\cite{Yosida:2012} higher-order spin-interactions such as the biquadratic or four-spin interaction are not required to model the magnetic structure. The spin-orbit interaction is then added in terms of perturbation theory in order to calculate the magneto-crystalline anisotropy energy and the DMI. 

The competition of the DMI with the exchange interaction leads to a long-period magnetic superstructure that can be inhomogeneous and since the generalized Bloch theorem is not applicable due to the presence of the spin-orbit interaction, a direct minimization of the total energy through an {\it ab initio} method is not attainable due to the large number of atoms involved in such a spiral. Therefore, it is convenient to derive out of the spin-model a micromagnetic model. Its solution is a spiraling magnetization density with an energy that depends on the pitch of the spiral. The energy as function of the pitch can then be compared to the total energy of {\it ab initio} calculations and permits thus the determination of the parameters entering the model from first-principles. With these parameters we determine from the micromagnetic model details of the magnetic structure.  

We find that the magnetic ground state of a monolayer Cr on W(110) has many similarities to the magnetic structure of Mn on W(110).\cite{Bode:07.1} For both we find a long-period homochiral left-rotating spin spiral driven by DMI on-top of a checkerboard-type c(2$\times$2) anti-ferromagnetic arrangement of magnetic moments between nearest neighbor atoms. The local anti-ferromagnetic structure is determined by the exchange interaction. In difference to the Mn system, the wave vector of the spiral in the Cr system is parallel to the  crystallographic $[001]$ direction of the surface and not along the [1$\bar{1}$0] direction. In a SP-STM experiment, a spin spiral along the $[001]$ direction results in a modulation of the magnetic contrast precisely along the measured direction, and thus our findings explain the long-period modulation in Cr/W(110) found by Santos {\it et al.}.\cite{Santos:08.1} Moreover, our calculations predict a period length of 14.3~nm, which is in excellent agreement to the experimentally observed value of $15.4\pm 1$~nm.\footnote{In an experimental STM image, vanishing magnetic contrast occurs at the nodes of a sinusoidal modulation and thus twice per period. This means, twice the experimental value of $7.7 \pm 0.5$~nm must be compared to the computed period length of 14.3~nm.} In contrast to Mn/W(110), we find a substantial inhomogeneity of the spin spiral, which increases the period length by 6\% as compared to the value of a homogeneous spin-spiral model.

The paper is organized as follows: in section II, the magnetic models are introduced in the form relevant to the Cr/W(110) system. The closest link to the {\it ab initio} results is provided by the classical spin model (section IIA). The Dzyaloshinskii-Moriya interaction is introduced in a formulation consistent with  a classical spin-model, which allows for a transparent analysis of possible magnetic structures by pure symmetry arguments. 
In section IIB, solutions of a micromagnetic model, which minimize the energy  containing the DMI are discussed, including their inhomogeneity and the criterion for the energetic stability. Section III describes the computational procedures taken  within DFT, from which our results for the Cr/W(110) system (section IV) are derived. In section V, the predicted ground state is discussed and compared to both, the experimental findings \cite{Santos:08.1} as well as the magnetic thin films of a monolayer Mn,\cite{Bode:07.1} a monolayer Fe\cite{Heide:09.1} and a doublelayer Fe on a W(110) substrate.\cite{Heide:08.1} A summary is provided in section VI.

\section{Magnetic Models}

\subsection{Spin Model}

The magnetic order and the thermodynamical properties of itinerant magnets can be frequently described by a classical atomistic spin model with parameters determined from first-principles.\cite{Lezaic:13} This holds true in particular if the $d$ electrons are fairly localized at the atomic site $i$, if the local magnetic moments are significantly large and their modulus shows only a minor dependence on the relative orientation of the magnetic moments, and if long-range exchange parameters encompassing the pair interaction between distant sites are taken into account.
Then, we can work with constant, localized magnetic moments, which are represented by a classical vector. 
Typically we work with unit vectors $\lbrace \vect{S}_i \rbrace$ $(i = 1 \ldots N)$, usually called `spin'. Expanding the energy of a spin system up to second order in $\vect{S}_i$, and keeping only the dominant terms, leads to two-site interactions and an on-site term, the spin model can be written as,\cite{Yosida:TheoryOfMagnetism, Udvardi:03.1}
\begin{equation}
  E = \sum_{i<j}{\left[ J_{ij}~{\vect S}_{i} \cdot {\vect S}_{j} + {\vect D}_{ij} \cdot \left({\vect S}_{i} \times {\vect S}_{j} \right) \right] } + \sum_{i}{{\vect S}_{i}^\T \, {\matr K}_{i} \, {\vect S}_{i}},
  \label{eq:model}
\end{equation}
where $J_{ij}$ is the exchange integral, ${\vect D}_{ij}$ is the Dzyaloshinskii vector and ${\matr K}_{i}$ is the on-site anisotropy term. The first term yields the classical isotropic Heisenberg model, the second term is the antisymmetric exchange or Dzyaloshinskii-Moriya interaction (DMI)\cite{Moriya:60.1, Dzyaloshinskii:57.1} and the third term is the magneto-crystalline anisotropy energy (MAE). 

The general solution of the Heisenberg model for a periodic lattice is a homogeneous spin spiral, which means that the angle $\varphi$ between two neighboring spins is constant, or linear combinations of these spirals that are symmetry related. The angle $\varphi_o$ describing the spin spiral with wave vector $\vect{q}_o$ that minimizes the Heisenberg energy depends on the set of exchange constants $\lbrace J_{ij} \rbrace$. This model contains two special solutions of collinearly aligned spins between nearest neighbor atoms, namely the ferromagnetic ($\varphi = 0$) and the anti-ferromagnetic phase ($\varphi=\pi$). However, the Heisenberg model only depends on $|\varphi|$ and thus two solutions to a given set $\left\lbrace J_{ij} \right\rbrace$ can be found minimizing the energy and representing two spin spirals differing just in their rotational sense. 

\begin{figure}[htb]
  \includegraphics[width=85mm]{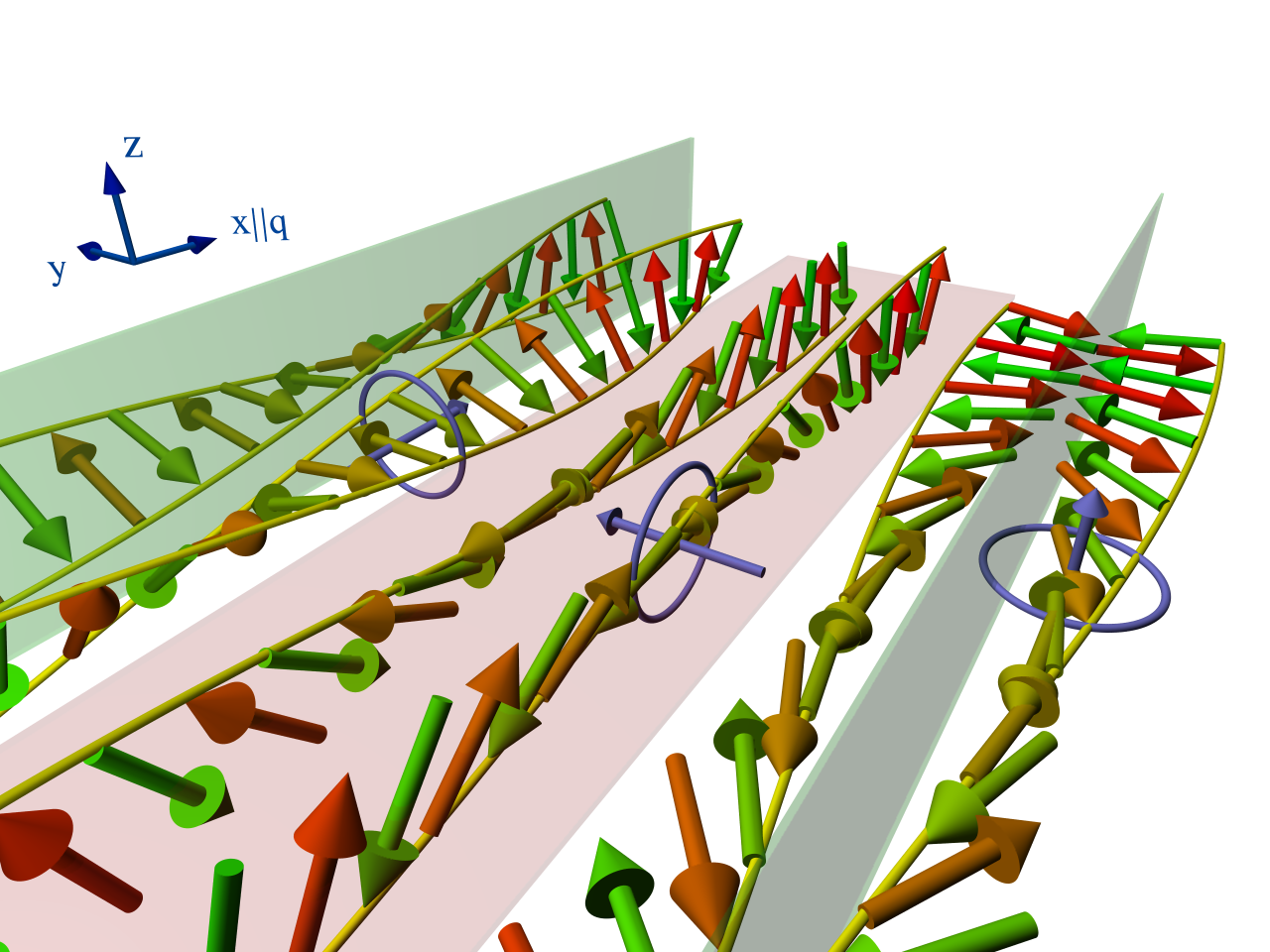}
  \caption{(Color online) Spin spirals with antiferromagnetic short-range order propagating with wave vector $\vect{q}$ along ${\uvect x}$, which is assumed to be a high-symmetry direction and thus parallel to a mirror plane (shaded planes) of the bulk lattice. We distinguish three pairs of spirals by the direction of their rotation axis $\vect{C}$ (indicated by a blue arrow) relative to the propagation direction and surface normal: one helical ($\vect{C} \parallel \vect{q}$) (left pair) and two cycloidal ($\vect{C} \perp \vect{q}$) spin spirals. The two spirals of each pair differ in their rotational sense, the left spiral being left handed. Only spirals with ${\vect C} \parallel {\uvect y}$ experience an influence of the DM interaction (see text).}
  \label{fig:dm-spirals}
\end{figure}

More generally, for a spiral the rotation axis of the magnetization of all atoms is the same and the spin spiral can be considered as mapping of the sites $i$, with $1\leq i \leq N$ to a unit circle in spin space $S^1$, where $N$ is the number of atoms in a spiral, {\it i.e.}\ the number of atoms in the spiral times the lattice constant defines the pitch of the spiral, $Na=\lambda$. If the mapping is (counter-)clockwise, that means the winding number of the circle is positive (negative) we speak of a right (left) rotating spiral with respect to the rotation axis. At this point it is convenient to introduce the vector chirality $\vect{C}=C \, \uvect{c}={\vect S}_i \times {\vect S}_{i+1}$. The direction of the vector chirality, $\uvect{c}$, acts as rotation axis. If $\uvect{c}$ can be restricted to the positive domain ($x \geq 0$,  $y \geq 0$, $z \geq 0$) of the lattice coordinates chosen such that $\uvect{x}$ is aligned parallel to the propagation vector $\vect{q}$ ({\it cf.}\ Fig.~\ref{fig:dm-spirals}), then it makes sense to speak about a right (left) rotating spiral if the value of the vector chirality $C$ or angle of rotation $\varphi$ is positive (negative), $C>0,\, \varphi>0$ ($C<0,\, \varphi<0$).

In contrast to the Heisenberg model, the DMI is sensitive to the rotational direction (due to ${\vect S}_i \times {\vect S}_j = - {\vect S}_j \times {\vect S}_i$), lifting this degeneracy and preferring a definite rotational sense depending on the sign of the Dzyaloshinskii-Moriya interaction denoted by $D$, of $\vect{D}= D \, \uvect{D}$ along the direction of the Dzyaloshinskii vector, $\uvect{D}$. The alignment of ${\vect D}$ with respect to the crystal lattice might be restricted due to symmetries, as outlined in Refs.~\onlinecite{Moriya:60.1} and \onlinecite{Crepieux:98.1}, respectively. For the case of spiraling magnetic structures propagating along a high symmetry line of a bcc(110) surface, ${\vect D}_{ij}$ must be perpendicular to the surface normal and to the propagation direction $\uvect{q}$ of the spiral. This can be understood from Fig.~\ref{fig:dm-spirals}: If the rotation axis ${\vect C}$ is parallel to $\vect q$ (left pair in Fig.~\ref{fig:dm-spirals}) or out-of-plane (right pair), the two spirals of different rotational sense are mirror images of each other with respect to a mirror plane of the lattice (shaded). Thus, the two spirals of different rotational sense of both pairs are of the same energy and as a consequence, the component of ${\vect D}_{ij}$ parallel to ${\vect C}$ must vanish. However, for ${\vect C}$ being perpendicular to both, the surface normal and ${\uvect q}$ (pair of spirals in the middle of Fig.~\ref{fig:dm-spirals}), the mirror plane connecting the two spirals of different rotational sense is parallel to the surface. This mirror symmetry is broken by the presence of the substrate. In this case we can conclude, that the direction ${\uvect D}$ must be perpendicular to both, the surface normal and ${\uvect q}$. From these considerations it is clear, that the DM interaction favors cycloidal spin spirals with a certain rotational sense, depending on the sign of ${D}$. This effect can be expected to be sizeable for systems with substrates of heavy elements of the periodic table (large atomic number $Z$) having a large spin-orbit interaction and breaking simultaneously the inversion symmetry. According to our sign convention of the Hamiltonian~\eqref{eq:model}, if $D<0$ ($D>0$) the energy can be lowered  by a right (left) rotating spiral.

\subsection{Micromagnetic Model}

In the previous chapter, we introduced the DMI in the language of a generalized Heisenberg model with localized magnetic moments on lattice sites. However, when the magnetic structure is slowly varying, meaning that the length scale on which the magnetization changes is large compared to the underlying atomic spacing, a continuum theory is beneficial providing analytical solutions. Since we are interested in spiral solutions we restrict ourselves to a one-dimensional (1D) micromagnetic model. Following Ref.~\onlinecite{Heide:11.1} a quasi-1D model is not only a good approximation for chains or narrow stripes, but also for monolayer-systems with a sufficiently anisotropic spin stiffness. In the 1D model, the magnetization is treated as a continuous vector field ${\vect m}(x)$ (with $|{\vect m}|=1$) instead of localized spins, and the Hamiltonian translates into an energy functional of the magnetization density and its spatial derivative ($\dot{\vect m} = \mathrm{d}{\vect m}/\dx$), \cite{Aharoni:Ferromagnetism}
\begin{equation}
  E[{\vect m}] = \frac{1}{\lambda} \int\limits_0^\lambda \dx \left[ \frac{A}{4\pi^2} ( \dot{\vect m} )^2 + \frac{\vect D}{2\pi} \cdot ( {\vect m} \times \dot{\vect m} ) + {\vect m}^\T {\matr K} {\vect m} \right],
  \label{eq:micmod}
\end{equation}
where $A$ is the spin stiffness corresponding to an effective isotropic exchange parameter originating from the set $\lbrace J_{ij} \rbrace$ for a 1D magnetic superstructure along $x$-direction, ${\vect D}$ is the (effective) Dzyaloshinskii-Moriya vector and ${\matr K}$ the anisotropy tensor of orthorhombic symmetry consistent with the Cr/W(110) system. $A$ and  ${\vect D}$ or more accurately $A_x$ and ${\vect D}_x$ depend on the direction of the 1D superstructure. $\lambda$ is the period length of the magnetic structure. The applicability of the micromagnetic model implies, that the exchange interactions $J_{ij}$ and ${\vect D}_{ij}$ decay sufficiently fast to justify the local character of Eq.~\eqref{eq:micmod}.

In the case of the bcc(110) surface, we choose our coordinate system such that the anisotropy tensor is diagonal, {\it i.e.}\
\begin{equation}\label{KdiagAniso}
{\matr K} = \mathrm{diag}(K_{1}, K_{2}, K_{3})\,.
\end{equation}
We further chose ${\vect D} = D \, {\uvect e}_3$ (${\uvect e}_3 = {\uvect y}$ in Fig.~\ref{fig:dm-spirals}) and define the directions $\uvect{e}_1$ and $\uvect{e}_2$ such that $K_2>K_1$, {\it i.e.}\ $\uvect{e}_1$ or $\uvect{e}_3$ is the easy axis. Finally we note,  in a micromagnetic model for ultrathin films the magnetostatic dipole-dipole interaction, which is a non-local contribution to the energy of the system, can be included into the local form of the anisotropy tensor (${\matr K} = {\matr K}^\mathrm{soc} + {\matr K}^\mathrm{dip}$). This is possible, because the dipole-dipole interaction between two magnetic moments decays on a length scale of a few nm, which is sufficiently smaller than the length scales involved in the micromagnetic model and it can thus be considered as local.\cite{Heide:08.1,Draaisma:88.1}

\begin{figure}[htb]
  \includegraphics[width=85mm]{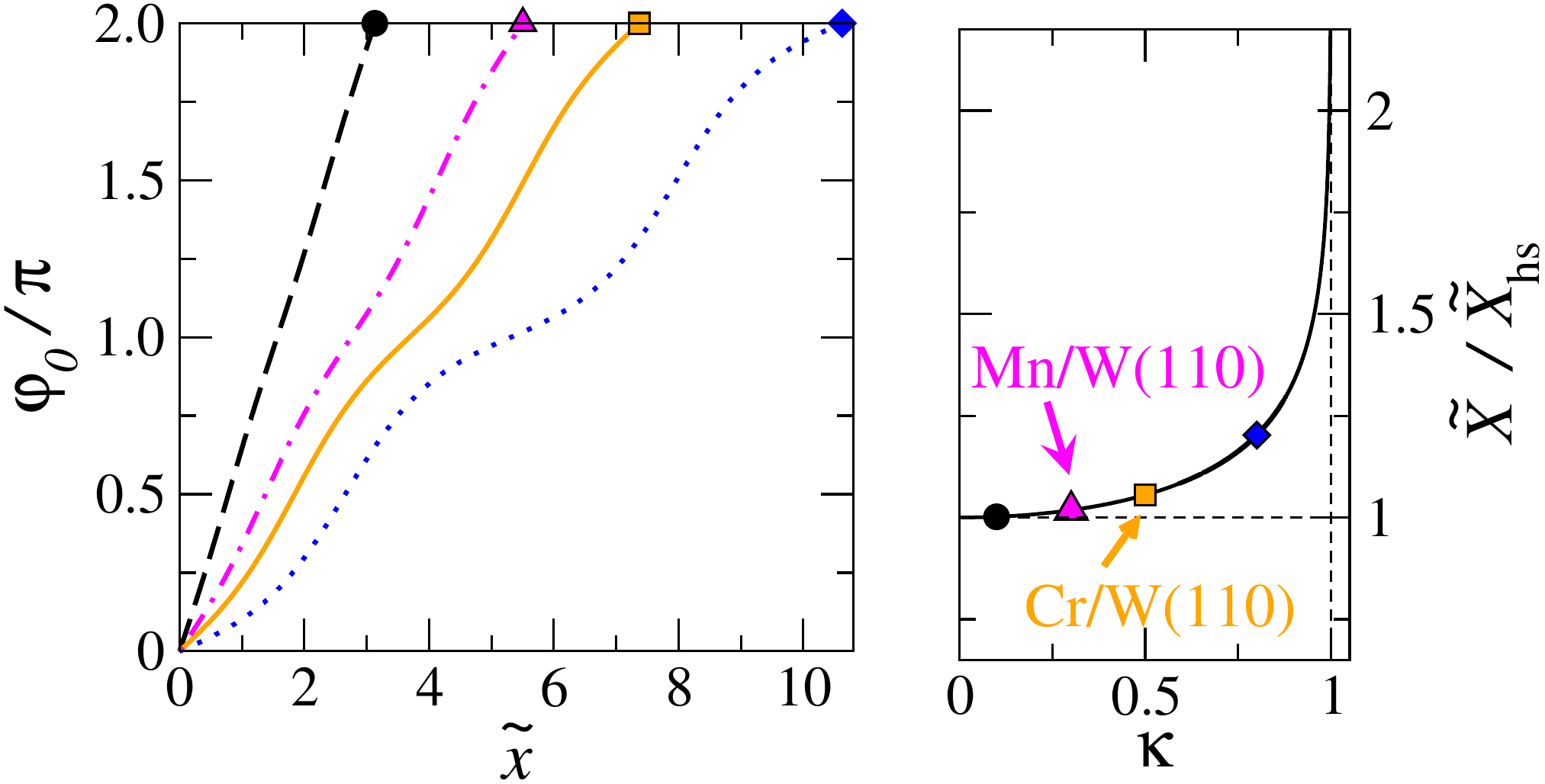}
  \caption{(Color online) Left: Profiles of inhomogeneous spin spirals with different inhomogeneities $\kappa=0.1$, 0.3, 0.5 and 0.8 (from left to right). The symbols denote the period length. Right: The period length diverges as $\kappa \nearrow 1 $, indicating a second order phase transition. $\tilde{X}_\mathrm{hs} = 4\pi \sqrt{AK}/|D|$ would be the optimal period length if the magnetic structure was constrained to homogeneous spirals. The values of $\kappa=0.3$ and 0.5 are obtained for the monolayers of Mn and Cr on W(110) ({\it cf.}\ Table~\ref{finaltablecrw}).}
  \label{fig:inhom}
\end{figure}

The model Eq.~\eqref{eq:micmod} comprises a rich phase diagram.\cite{Heide:11.1} Depending on the model parameters $A$, ${\vect D}$ and ${\matr K}$, four magnetic phases can be distinguished: a phase with collinear magnetization perpendicular or parallel to ${\vect D}$, respectively, a phase with non-collinear magnetization confined to a plane perpendicular to ${\vect D}$ and a truly 3-dimensional magnetic structure. The type of spin spirals crucially depends on the direction of the easy axis ({\it i.e.}\ the direction of lowest energy). The truly three-dimensional ground state only exists for systems with a small difference in the anisotropy energies between two directions (easy-plane anisotropy), of which one must be the direction of ${\vect D}$. However, the corresponding section of the phase space is rather small and we restrict our further analysis to magnetization densities which are not truly three-dimensional, but confined to a plane perpendicular to ${\vect D}$ or $\uvect{e}_3 = \uvect{y}$, respectively. In other words, the rotation axis is parallel to $\vect{D}$ and the ground-state energy of the system can get lowered by the DMI. For such a magnetic structure, the magnetization direction has only one degree of freedom and thus can be described by a single angle $\varphi = \varphi(x)$, \textit{i.e.}
\begin{equation}\label{mag_flat_ss}
  {\vect m}(x) = \cos(\varphi)\,{\uvect e}_1 + \sin(\varphi)\,{\uvect e}_2\,,
\end{equation}
and the energy functional Eq.~\eqref{eq:micmod} can be (up to a constant term) written as
\begin{equation}\label{micmod:flatspiral}
  \tilde{E}[\varphi] = \frac{1}{\tilde{X}} \int_0^{\tilde{X}} \dd \tilde{x} \left[ \left( \frac{\dd \varphi}{\dd \tilde{x}} \right)^2 + \tilde{D} \, \frac{\dd \varphi}{\dd \tilde{x}} + \sin^2{\varphi} \right]
\end{equation}
with reduced parameters
\begin{eqnarray}\label{micmod:parameters}
 \tilde{D} &=& \frac{D}{\sqrt{A\,K}} ,~~~ \tilde{E} = \frac{E}{K} ,~~~ K = K_2-K_1, \\
 \tilde{x} &=& \frac{2\pi \, x}{\sqrt{A/K}} ,~~~ \tilde{X} = \frac{2\pi \, \lambda}{\sqrt{A/K}} , \nonumber
\end{eqnarray}
which is dependent on one effective parameter $\tilde{D}$.
The magnetization profile\cite{Dzyaloshinskii:65.2}
\begin{equation}
  \varphi_0(\tilde{x}) = (-\sign \tilde{D}) ~ \am\left( (\tilde{x}-\tilde{X}/4)/\epsilon, \epsilon \right)\label{micmod:profile}
\end{equation}
minimizes the total energy and the ground-state energy can be written as
\begin{equation}
  \tilde{E}_\mathrm{min} := \tilde{E}[\varphi_0] = 1 - \frac{1}{\epsilon^2}  ~.\label{micmod:minimalenergy}
\end{equation}
$\am(x,\epsilon)$ is the Jacobi elliptic amplitude function and $\epsilon$ corresponds to the inhomogeneity of the spiral. The latter and the reduced period length $\tilde{X}$ are determined via the complete elliptic integrals of first and second kind, $\mathrm{K}(\epsilon)$ and $\mathrm{E}(\epsilon)$,\footnote{The complete elliptic integrals of first and second kind are defined as $K(\epsilon)=\int_0^{\pi/2}{\mathrm{d}\phi \frac{1}{\sqrt{ 1-\epsilon^2 \, \sin^2 \phi }}}$ and $E(\epsilon) = \int_0^{\pi/2}{\mathrm{d}\phi \sqrt{ 1-\epsilon^2 \, \sin^2 \phi }}$}respectively,
\begin{equation}
\mathrm{E}(\epsilon) / \epsilon  = \pi | \tilde{D} | / 4 ~~ \mathrm{and} ~~  \tilde{X} = 4\epsilon \, \mathrm{K}(\epsilon)\,.
\end{equation}
Thus, the inhomogeneity parameter $\epsilon$ is defined implicitly and depends on the model parameter $\tilde{D}$. Therefore, it is convenient to introduce another dimensionless measure for the inhomogeneity, {\it i.e.}\
\begin{equation}
  \kappa = \left( \frac{\epsilon}{ \mathrm{E}(\epsilon)} \right)^2 = \left(\frac{4}{\pi}\frac{1}{\tilde{D}}\right)^2=\left(\frac{4}{\pi}\right)^2\frac{AK}{D^2}\,,~~ 0 \leq \kappa < 1.\label{kappa:definition}
\end{equation}
For the parameter set $\kappa\in [0,1[$  a periodic spin spiral takes the lowest energy. The spiral is homogeneous for infinitesimally small positive $\kappa$ ($\kappa \searrow 0$), meaning that the slope of the profile, $\mathrm{d}\varphi_0/\mathrm{d}\tilde{x}$, is constant, and becomes maximally inhomogeneous for $\kappa \nearrow 1$. In the latter case, the magnetization rotates slowly when it points in direction of the easy axis ${\uvect e}_1$ ({\it e.g.}\ regions with $\varphi=0$ and $\varphi=\pi$, {\it cf.}\ Eq.~\eqref{mag_flat_ss}) and rotates very fast over the harder axis ${\uvect e}_2$ ({\it e.g.}\ $\varphi = \pi/2$ and $\varphi=3\pi/2$, respectively). This represents two collinearly magnetized domains of opposite orientation with a domain wall in between and explains the chiral domain walls found in the  the stripes of Fe doublelayers on the stepped W(110) surface.\cite{Heide:08.1} Profiles of spin spirals according to Eq.~\eqref{micmod:profile} are shown in Fig.~\ref{fig:inhom}. Additionally, the pitch $\lambda$ of an inhomogeneous spiral relative to the pitch of a spin spiral constraint to a homogeneous rotation is shown, which diverges as $\kappa \nearrow 1 (D \searrow \frac{4}{\pi}\sqrt{AK})$ and undergoes a second order phase transition into the collinear phase with magnetization perpendicular to ${\vect D}$ for $\kappa>1$.\ \cite{Dzyaloshinskii:65.2,Izyumov:84.1,Heide:11.1}

We finally elaborate on a simplified micomagnetic model constrained to homogeneous spin spirals, for which the angle of rotation, $\varphi_\mathrm{hs} =2\pi/\lambda\cdot x$ increases linearly as function of $x$, and thus $\dd \varphi/\dd x = \mathrm{const.}$. The energy functional Eq.~\eqref{micmod:flatspiral} turns into a function of the period length $\lambda = 2\pi (\dd \varphi/\dd x)^{-1}$,
\begin{equation}  \label{dispersion}
  E(\lambda) = A\,/\lambda^2 + D\,/\lambda + K/2~.
\end{equation}
This energy becomes minimal for a period length $\lambda_\mathrm{hs} = -2A/D$ and takes the value
\begin{equation}  \label{hom_ss:minenergy}
  E_\mathrm{hs} = \frac{K}{2}\left(1-\frac{1}{2}\frac{D^2}{AK}\right).
\end{equation}
In this model constrained to homogeneous spin spirals, the ground state will  be non-collinear only if $E_\mathrm{hs}<0$, which translates to the condition $D^2 > 2AK$. Inhomogeneous spirals become unstable against the collinear state for a 20\%  smaller $D$, $D^2 \geq 16/\pi^2 AK$ ({\it cf.}\ Eq.~\eqref{kappa:definition}).

\section{Computational Methodology}

We investigate the magnetic structure from first-principles based on non-collinear spin-density functional theory. The short-ranged magnetic structure is investigated by total-energy calculations between different collinear magnetic states in the c(2$\times$2) unit cell. The DMI and the MAE are a direct consequence of the spin-orbit interaction, and thus require relativistic calculations, while the relativistic calculation alter the exchange parameter only slightly. The long-period magnetic ground state is then determined on the basis of the micromagnetic model~\eqref{micmod:flatspiral} with model parameters $A$, $D$, and $K$, that we determine from first-principles. According to Eq.~\eqref{dispersion}, the model parameters $A$ and $D$ can be obtained by parabolic and linear fits of the total-energy results for homogeneous spin spirals with different period lengths $\lambda$. For the antisymmetric contribution, spin-spiral calculations including spin-orbit coupling must be performed.\cite{Heide:09.1}  The magnetic anisotropy tensor, $\matr{K}$, is obtained from relativistic collinear DFT calculations carrying out self-consistent calculations or employing the force theorem. The contribution of the dipole-dipole interaction to the magnetic anisotropy tensor can be neglected in the case of a Cr monolayer, because it is of minor importance for this antiferromagnet. The exact details of the choice and application of the methods and their limits, are discussed in detail below.

All calculations make use of the non-collinear formulation of the full-potential linearized augmented planewave (FLAPW) method\cite{Kurz:04.1} and the relativistic extension\cite{Heide:09.1} in the two-dimensional slab  geometry\cite{Krakauer:79.1, Wimmer:81.1} describing a film perfectly embedded into two semi-infinite vacua, all implemented in the {\sc Fleur}-code.\cite{FLEUR}

\subsection{Magneto-crystalline Anisotropy Energy and Force Theorem}
\label{chap:method.forcetheorem}

Spin-orbit coupling can be derived as correction term to the ordinary Schr\"odinger equation from a $1/c$-expansion of the fully relativistic Dirac equation. In a spherical potential, the spin-orbit operator takes the form $\mathcal{H}_\mathrm{so} = \xi (r) \, \boldsymbol{\sigma} \cdot \vect{L}$, where $\xi \sim {r}^{-1} \, \mathrm{d}V/\mathrm{d}r$. Thus, spin-orbit coupling is largest in the vicinity of the nucleus, where the potential has the 1/r singularity and is indeed nearly spherical symmetric. Thus, we can safely approximate the presence of the SOC operator to the muffin-tin spheres, {\it i.e.}\
\begin{equation}\label{SOCoperator}
  \mathcal{H}_\mathrm{so} = \sum_{\mu} 
  \xi (r^{\mu}) \, \boldsymbol{\sigma} \cdot {\vect L}^{\mu} \, ,
\end{equation}
where ${\vect r}^{\mu} = {\vect r} - {\vect R}^{\mu}$ and $|{\vect r^{\mu}}| < R^{\mu}_{\mathrm{MT}}$. ${\vect R}^{\mu}$ is the center and $R^\mu_\mathrm{MT}$ is the radius of the $\mu$th muffin-tin sphere in the unit cell, and $\mu$ runs over all atoms in the unit cell.

In order to calculate the MAE or $\matr{K}$, respectively, we calculate the total energy for magnetic moments aligned collinearly ({\it i.e.}\ antiferromagnetically for Cr) along all three high-symmetry directions of the system, {\it i.e.}\ along $[110]$, $[001]$ and $[1\overline{1}0]$. Instead of performing self-consistent relativistic calculations for all three directions, it is a reasonable approach to calculate the charge and magnetization density, $n_0$ and $\vect{m}_0 = m_0 \, \uvect{m}_0$ respectively, self-consistently with spin-orbit coupling but only along one direction, {\it e.g.}\ $\uvect{e}_1$, and then solving the secular equation only once for each remaining magnetization direction $\uvect{e}_k$ ($k=2,3$). The force theorem works reliably if the charge and magnetization density change little after rotation of the quantization axis, {\it i.e.}\ if  $\delta n = n_0(\uvect{e}_k) - n_0(\uvect{e}_1)$ and $\delta m = m_0(\uvect{e}_k) - m_0(\uvect{e}_1)$ is small. In W the strength of SOC is fairly large and influences the electronic structure considerably. This makes it important to include SOC in the self-consistent calculation of $(n_0, \, \vect{m}_0)$, but the difference in the magnetization directions $\vect{m}_0$ can be regarded as perturbation ({\it cf.}\ Sec.~\ref{chap:results.anisotropy}).

According to Andersen's force theorem (FT),\cite{Mackintosh:80.1, Oswald:85.1, Liechtenstein:87.1} the change in total energy $\delta E$ due to this perturbation can be approximated by a summation over all occupied (occ.) states ($\nu$ being the band index, and $\vect{k}$ is the Bloch vector),
\begin{equation}\label{ForceTheoremEnergy}
E_\mathrm{MAE}= \delta E \approx \sum_{\vect{k}\nu}^{\mathrm {occ.}}{\varepsilon^{\mathrm{FT}}_{\vect{k}\nu}}(\uvect{e}_k) - \sum_{\vect{k}\nu}^{\mathrm {occ.}}{\varepsilon^{0}_{\vect{k}\nu}}(\uvect{e}_1)\,,
\end{equation}
which is equal the magneto-crystalline anisotropy energy measured with respect to the direction $\uvect{e}_1$. $\{ \varepsilon^{\mathrm{FT}}_{\vect{k}\nu} \}$ and $\{ \varepsilon^{0}_{\vect{k}\nu} \}$ is the spectrum of the perturbed and unperturbed Hamiltonian, both constructed from the unperturbed electron density and the selected magnetization direction, 
\begin{eqnarray}\label{ForceTheoremHamiltonian}
  \left( \mathcal{H}_0 + \delta \mathcal{H} \right) [n_0,\, m_0,\, \uvect{e}_1] \, \psi^\mathrm{FT}_{\vect{k}\nu} = \varepsilon^{\mathrm{FT}}_{\vect{k}\nu}(\uvect{e}_k) \, \psi^\mathrm{FT}_{\vect{k}\nu} \label{eq:secular_forceMCA1}\\
  \mathcal{H}_0 [n_0,\, m_0,\, \uvect{e}_1] \, \psi^{0}_{\vect{k}\nu} = \varepsilon^{0}_{\vect{k}\nu}(\uvect{e}_1) \, \psi^{0}_{\vect{k}\nu} \,.
\end{eqnarray}
In this way we can construct $\matr{K}^\mathrm{soc}$ from the same self-consistent charge density solving the secular equation \eqref{eq:secular_forceMCA1} only three times.

\subsection{Spin Stiffness and Generalized Bloch Theorem}
To obtain the spin stiffness $A$ along a certain direction, we calculate homogeneous spin spirals with wave vector $\vect{q}$ along this direction. The spin-stiffness stands for the exchange interaction caused by the Coulomb interaction between electrons and the Pauli principle. Since the Coulomb interaction is of much larger energy scale than the spin-orbit interaction, $A$ can be calculated to a good approximation by means of the scalar-relativistic approximation, \textit{i.e.}\ neglecting spin-orbit coupling. Then, we can choose an arbitrary orientation of the rotation axis of the spin spiral relative to the lattice coordinates. For convenience we chose the $\hat{z}$-direction, thus the spin spiral rotates in the plane spanned by ${\uvect e}_1 = \hat{x}$ and ${\uvect e}_2 = \hat{y}$. The rotation angle $\varphi$ of the direction of the spin-moment within each muffin-tin sphere varies from atom to atom as $\varphi(\boldsymbol{\tau}^{\mu}+\vect{R}^n) = \varphi(\boldsymbol{\tau}^{\mu}) + \vect{q} \cdot \vect{R}^n$, where $\boldsymbol{\tau}^{\mu}$ denotes the position of the $\mu$th basis atom within the unit cell, and $\vect{R}^n$ is a lattice vector. The use of the generalized Bloch theorem \cite{Sandratskii:91.1} allows us to perform calculations for incommensurable spin spirals using the chemical unit cell, because the eigenstates of a Schr\"odinger-type Hamiltonian with this type of magnetic symmetry can be decomposed into a ${\vect q}$-dependent phase factor (corresponding to a spin rotation of angle $\alpha = \vect{q} \cdot \vect{r}$ around the ${\hat z}$-axis by means of matrix $\matr{U}(\alpha)$) and an ordinary Bloch function with lattice-periodic Bloch factors $u^{\sigma}_{\vect{k}\nu}({\vect r}) = u^{\sigma}_{\vect{k}\nu}({\vect r} + \vect{R}^n)$,
\begin{equation}\label{GenBlochTheorem}
  \psi_{\vect{k}\nu} ({\vect r}|{\vect q}) = \expo{ - i {\matr \sigma}_z \, {\vect q} \cdot {\vect r}/2 } ~ \expo{i {\vect k} \cdot {\vect r}}  \, \left( \begin{array}{c} u^\uparrow_{\vect{k}\nu} ({\vect r}) \\  u^\downarrow_{\vect{k}\nu} ({\vect r}) \end{array} \right) \,,
\end{equation}
where $\sigma_z$ is the $2\times 2$ Pauli matrix, and the first exponential function on the right hand side corresponds to the spin-rotation matrix $\matr{U}(\vect{q} \cdot \vect{r})$.

The resulting energy dispersion, \textit{i.e.}\ the DFT total energy $E_\mathrm{SS}$ of the spin spiral state for a given wave vector $\vect{q}$, is symmetric around the ferromagnetic and antiferromagnetic state, and the spin stiffness $A$ of Eq.~\eqref{dispersion} can be obtained by a quadratic fit of the total energy, $E_\mathrm{SS}(\vect{q}) \propto A \, \abs{\vect{q}}^2$ around the FM or $E_\mathrm{SS}(\vect{q}) \propto A \, \abs{\vect{q} - \vect{q}_\mathrm{AFM}}^2$ around the AFM.

For small changes $\delta \vect{q}$ around a fixed wave vector $\vect{q}_0$ the change of the total energy $\delta E_\mathrm{SS} = E_\mathrm{SS}(\vect{q}_0 + \delta \vect{q})-E_\mathrm{SS}(\vect{q}_0)$ can be calculated  employing again the force theorem. The deviation in the spin-spiral vector, $\delta \vect{q}$, results in a perturbation to the Hamiltonian that depends on the self-consistent electron and magnetization density, $(n_0,\, {\vect m}_0)$, obtained for the fixed wave vector $\vect{q}_0$. In our case this allows the calculation of the spin-stiffness by performing self-consistent scalar-relativistic calculations only for the antiferromagnetic state $\vect{q}_0 = \vect{q}_\mathrm{AFM}$ and subsequent diagonalizations of the Kohn-Sham-Hamiltonian once for each $\vect{q}$.

\subsection{DMI from First Order Perturbation Theory} \label{sec:meth:DMI}

In leading order the antisymmetric Dzyaloshinskii-Moriya interaction depends linearly on the spin-orbit coupling strength, $\xi$. Since the spin-orbit coupling is small relative to the kinetic energy or the different potential energies of the Hamiltonian, as described in Ref.~\onlinecite{Heide:09.1}, first order perturbation theory is a convenient way to calculate the DMI to a good approximation.
Thus, we calculate the matrix elements
\begin{equation}\label{FstO_expctval}
  \delta{\epsilon}_{\vect{k}\nu} (\vect{q})= \langle \matr{U}_\mathrm{g}\psi_{\vect{k}\nu}(\vect{q}) | \mathcal{H}_\mathrm{so} | \matr{U}_\mathrm{g} \psi_{\vect{k}\nu}(\vect{q}) \rangle
\end{equation}
of the operator Eq.~\eqref{SOCoperator} with the states $\ket{\psi_{\vect{k}\nu}(\vect{q})}$ from Eq.~\eqref{GenBlochTheorem}. However, in the derivation of Eq.~\eqref{GenBlochTheorem} explicit use was made by the fact that the non-relativistic Hamiltonian is invariant under a global spin-rotation, and thus the rotation axis $\vect{C}$ of the spin spiral was chosen along the $z$ axis. The spin-orbit coupling breaks this invariance and the states $\psi_{\vect{k}\nu}(\vect{q})$ require a rotation by a global spin rotation $\matr{U}_\mathrm{g}$, which brings $\vect{C}$ into the desired direction.

Summing up all energy shifts from occupied states yields the DMI energy in first order perturbation theory,
\begin{equation}
  E_\mathrm{DMI}(\vect{q}) = \sum_{\vect{k}\nu}{ n_{\vect{k}\nu}({\vect{q}}) ~ \delta{\epsilon}_{\vect{k}\nu} (\vect{q}) } ~,
\end{equation}
$n_{\vect{k}\nu}({\vect{q}})$ being the occupation numbers that correspond to state $\vert\psi_{\vect{k}\nu}(\vect{q}) \rangle$, and we can extract $D$ by a linear fit $E_\mathrm{DMI}(\vect{q}) \propto D \, q$.

Each Kohn-Sham orbital exhibits the symmetry $\delta{\epsilon}_{\vect{k}\nu} (-\vect{q}) = - \delta{\epsilon}_{\vect{k}\nu} ( \vect{q})$, and because the occupation numbers of the states, do not depend on the sign of $\vect{q}$, also the sum inherits this anti-symmetry and only spin spirals of one rotational sense have to be calculated.


In a spin spiral, the angle between the magnetization and the crystal lattice varies from atom to atom along the spin-spiral vector. Thus, due to the spin-orbit interaction each atom contributes differently to the spin-orbit induced energy although the atoms are chemically equivalent. Therefore, the summation in Eq.~\eqref{SOCoperator} is to be understood over all atoms in the magnetic supercell. For first order perturbation theory, however, one can still restrict the integration contained in Eq.~\eqref{FstO_expctval} to the chemical unit cell as is shown in Appendix~\ref{sec:appendix}. This makes the scheme computationally very efficient.

Moreover, the site decomposition of the SOC operator in Eq.~\eqref{SOCoperator} allows us to obtain a layer-resolved DMI energy $E_\mathrm{DMI}^{\mu}(\vect{q})$, where $\mu$ labels the atom in the unit cell.

\subsection{Computational Details}
The calculations make use of two different approximations to the unknown exchange-correlation functionals in DFT. One is the generalized gradient approximation (GGA) PBE\cite{Perdew:96.1} used for the structural optimization and the other one a local density approximation (LDA)\cite{Moruzzi:78} for the determination  of the energetics between the different magnetic structures.
Also two different structural models are taken for the two types of calculations: For the structural optimization the Cr film on the W(110) substrate is modeled by a slab of 7 layers of W covered with a Cr monolayer on both sides, and 7 layers of W covered with a Cr monolayer on only one side are chosen for the analysis of the magnetic states. The muffin-tin radii were chosen to $2.3~\mathrm{a.u.}$ for Cr and $2.5 \, \mathrm{a.u.}$ for W. The APW basis functions are expanded up to a wave vector of $k_\mathrm{max} = 3.8~\mathrm{a.u.}^{-1}$ and in the muffin-tin spheres basis functions including spherical harmonics up to $\ell_\mathrm{max} = 8$ were taken. The full two-dimensional Brillouin zone corresponding to the p(1$\times$1) unit cell was sampled by 4608 $\vect{k}$-points for the calculation of the DM vector, and up to 10368 $\vect{k}$-points were used for the calculation of the spin stiffness. The MAE was calculated with up to 2304 $\vect{k}$-points in the full Brillouin zone corresponding to the c(2$\times$2) unit cell ({\it cf.}\ Sec.~\ref{chap:results.anisotropy}), respectively.
All calculations of the bcc-structured W-substrate are carried out with the GGA bulk lattice constant, $a=6.03~\mathrm{a.u.}$, and the Cr-W distance relaxed to $3.90 \, \mathrm{a.u.}$. This corresponds to an inwards relaxation of $8.6\%$, consistent with experimental ($8.0 \pm 0.7 \%$) and computational ($8.5\%$) findings in Ref.~\onlinecite{Santos:08.1}. The structural relaxation was done assuming an antiferromagnetic ground state and neglecting SOC.

\begin{figure}[tbh]
  \includegraphics[width=50mm]{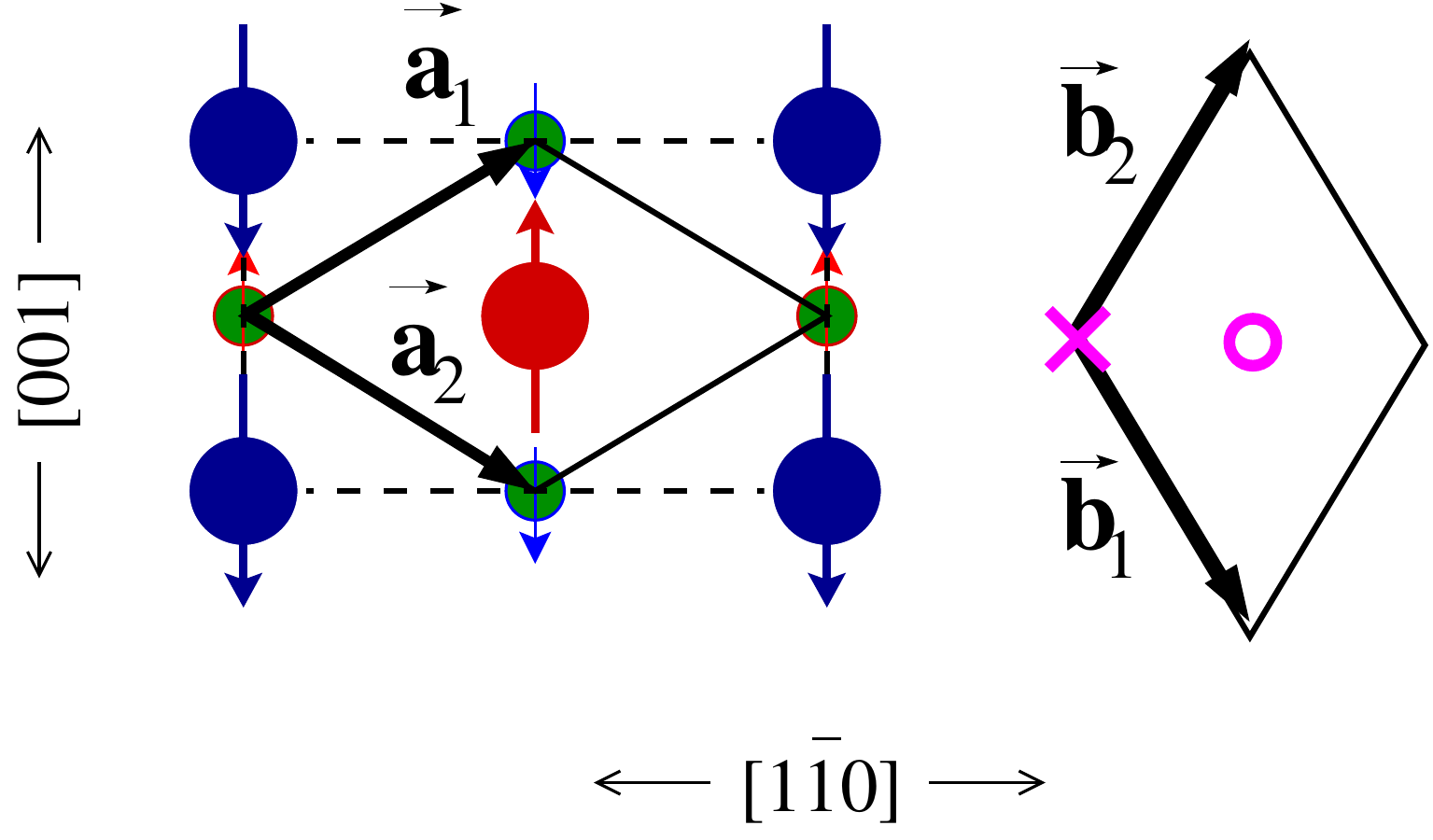}
  \caption{(Color online) Left: Arrangement of atoms in Cr/W(110). Big circles represent Cr atoms with magnetic moments pointing up and down, respectively, and small circles indicate the positions of the W atoms of the first layer of the substrate. The dashed line represents the magnetic c(2$\times$2) unit cell in a collinear calculation, and $\vect{a}_1=\frac{1}{2}(\sqrt{2}a,a)$ and $\vect{a}_2=\frac{1}{2}(\sqrt{2}a,-a)$ indicate the p(1$\times$1) unit cell of the chemical lattice. Right: Reciprocal lattice vectors $\vect{b}_1$ and $\vect{b}_2$ for the chemical unit cell. The cross and circle represents those $\vect{q}$-vectors associated with the FM and AFM state, respectively.}
  \label{fig:unitcell}
\end{figure}

Most of the calculations are carried out in the two-dimensional chemical p(1$\times$1) unit cell, which comprises only one surface atom, or one atom per layer, respectively (see Fig.~\ref{fig:unitcell}). Also the antiferromagnetic state can be described in the chemical unit cell as long as spin-orbit coupling is neglected, by making use of the generalized Bloch theorem and choosing $\vect{q}_\mathrm{AFM} = ( \vect{b}_1 + \vect{b}_2 ) / 2$. $\vect{b}_1$ and $\vect{b}_2$ are the reciprocal lattice vectors (see Fig.~\ref{fig:unitcell}). For the calculation of the structural relaxation, the short range magnetic order and the MAE, however, a c2($\times$2) supercell comprising two surface atoms was used.

By choosing a deviation $\delta \vect{q}$ from $\vect{q}_\mathrm{AFM}$, we induce a spin spiral with antiferromagnetic short-range order of period length $\lambda = 2\pi/|\delta \vect{q}|$. To be more precise, $\delta \vect{q} = q' (\vect{b}_1 + \vect{b}_2)$ induces a spiral propagating along the $[1\overline{1}0]$ direction with the pitch $\lambda = a/(\sqrt{2}q')$ and $\delta \vect{q} = q' (\vect{b}_1 - \vect{b}_2)$ induces a spiral along $[001]$ with $\lambda = a/(2q')$, where $a$ is the lattice constant of W. Thus, the period length of the AFM ($q'=0$) is defined to be infinite in the notion of the magnetic supercell. The sign of $q'$ determines the rotational sense of the spiral, $\mathrm{sign}(q') = \mathrm{sign}(\lambda)<0$ being a left-handed spin spiral.

\section{Results}

Total-energy calculations of Cr/W(110) neglecting SOC prove that the checkerboard-type AFM state (see Fig.~\ref{fig:unitcell}) is preferred over the FM one by 200~meV per Cr atom, in accordance with Ref.~\onlinecite{Santos:08.1}. This is a large energy scale, and therefore, all subsequent calculations take the AFM state as starting point. The magnetic moment of Cr is then $2.41~\mathrm{\mu_B}$ and the induced W moments at the interface are $0.2~\mathrm{\mu_B}$. They couple antiferromagnetically to the nearest neighbor Cr atoms. As a result, the induced W atoms are also arranged in a checkerboard c(2$\times$2) antiferromagnetic order. Even if the induced $W$ moments would tend to couple ferromagnetically among each other, the corresponding interaction energy would be much smaller compared to the interaction energy between the W and Cr moments due to the smallness of the W moments. Since the W moments are only induced by Cr and the W-W interaction is negligible, it is sufficient to work with an effective spin-model in which only the Cr moments enter.\cite{Lezaic:13}	

\subsection{Magnetic Anisotropy}
\label{chap:results.anisotropy}
The magnetic anisotropy energy (MAE) has two origins, namely the spin-orbit coupling of the electrons and the shape anisotropy, which results from the magnetostatic dipole-dipole interaction between the atomic magnetic moments of a sample. However, the latter is small in 2D systems and basically negligible  due to the antiferromagnetic alignment of neighboring magnetic moments for Cr on W(110). Thus, the dipolar interaction will be neglected in the following. The contribution from SOC is obtained by electronic structure calculations: three approaches were used, namely (I) self-consistent calculations including the SOC operator for magnetic moments pointing into different directions and (II) using the magnetic force theorem, {\it i.e.}\ Eq.~\eqref{ForceTheoremHamiltonian}, as described in Sec.~\ref{chap:method.forcetheorem}. For the force theorem, the secular equation was constructed from the converged charge density $(n_0,\vect{m}_0)$ with the magnetization pointing in the $[001]$- or $[110]$-direction (labeled IIa or IIb in Table~\ref{TableMAE}), respectively. Additionally, we repeat the calculations of approach (IIa), but for a film thickness increased by one more W layer (approach IIIa). All approaches are evaluated for the antiferromagnetic state in the c(2$\times$2) unit cell with two atoms per layer using the collinear version of the  FLAPW code. Please note, that the corresponding Brillouin zone is smaller by a factor 2 as compared to calculations in the p(1$\times$1) unit cell (Fig.~\ref{fig:unitcell}) and correspondingly for a given number of $\vect{k}$-points, the $\vect{k}$-point mesh is twice as dense as in a calculation with only one atom per layer in the unit cell.

\begin{table}
 \begin{ruledtabular}
  \begin{tabular}{llllll}
     & \multicolumn{1}{c}{$K^\mathrm{soc}_{[001]}$} & \multicolumn{1}{c}{$K^\mathrm{soc}_{[1\bar{1}0]}$} \\[1.2ex] \hline
approach (I)    & 0.86  & 1.28 \\
approach (IIa)  & 0.92  & 1.18 \\
approach (IIb)  & 0.93  & 1.20 \\
approach (IIIa) & 1.06  & 1.21 \\
  \end{tabular}
 \end{ruledtabular}
 \caption{Values for the magneto-crystalline anisotropy energy (MAE) with respect to the easy axis (determined to be out-of-plane, {\it i.e.}\ along $[110]$) in meV per magnetic atom. See text for a description of the approaches.}
 \label{TableMAE}
\end{table}

We determined the $[110]$-direction (out-of-plane) to be the easy axis and the in-plane $[1\bar{1}0]$-direction to be the hard axis. The approaches need up to 2304 $\vect{k}$-points to yield converged results. The values (see Table~\ref{TableMAE}) between approaches (I) and (II) agree within $15\%$. For the approach (II), the values for the MAE are independent on the direction of $\vect{m}_0$ (compare IIa and IIb). The influence of the number of layers of the a film (consisting of $7$ and $8$ W-layers; compare  IIa and IIIa) also yield consistent results. We estimate the final anisotropy constants to be $K^\mathrm{soc}_{[001]}= 0.9~\mathrm{meV}$ and $K^\mathrm{soc}_{[1\bar{1}0]}= 1.2~\mathrm{meV}$ with uncertainties of about $0.1$~meV.

\begin{table}
 \begin{ruledtabular}
  \begin{tabular}{llllll}
     & \multicolumn{1}{c}{1 (Cr)} & \multicolumn{1}{c}{2 (W)} & \multicolumn{1}{c}{3 (W)} & \multicolumn{1}{c}{4 (W)} & \multicolumn{1}{c}{5 (W)} \\[1.2ex] \hline
   $\mu_\mathrm{S} \, [\mu_\mathrm{B}]$        & $\phantom{-}2.42$ & $\phantom{-}0.21$ & $\phantom{-}0.04$             & $\phantom{-}0.01$ & $\phantom{-}0.01$ \\[1.2ex]
   $\mu_\mathrm{L} \, [10^{-2}\mu_\mathrm{B}]$ \\
   \hspace{1.0em} $\vect{m}\parallel [1\overline{1}0]$ & $-1.0\phantom{0}$ & $-0.5\phantom{0}$ & $-1.0\phantom{0}$ & $-0.1\phantom{0}$ & $\phantom{-}0.0\phantom{0}$ \\
   \hspace{1.0em} $\vect{m}\parallel [001]$ & $-2.4\phantom{0}$ & $-1.6\phantom{0}$ & $-0.3\phantom{0}$ & $-0.3\phantom{0}$ & $-0.3\phantom{0}$ \\
   \hspace{1.0em} $\vect{m}\parallel [110]$ & $-2.0\phantom{0}$ & $-0.8\phantom{0}$ & $-0.1\phantom{0}$ & $\phantom{-}0.0\phantom{0}$ & $-0.1\phantom{0}$ \\
  \end{tabular}
 \end{ruledtabular}
 \caption{Spin ($\mu_\mathrm{S}$) and orbital ($\mu_\mathrm{L}$) magnetic moments for the first five layers. The orbital moments are given for different directions of the magnetization $\vect{m}$.}
 \label{TableMagMom}
\end{table}

In Table \ref{TableMagMom}, we list the spin- and orbital magnetic moments of the first five layers in the AFM sate of the thin film as calculated self-consistently including SOC. Spin- and orbital moments of all atoms couple antiferromagnetically consistent with Hund's third rule for less than half filled $d$-shells. The largest orbital moments we find for Cr, but the orbital moments of W are of the same magnitude and not one or two orders of magnitude smaller as in the case of the spin moments. Since spin-orbit coupling is crucial for the development of an orbital moment, the value of $\mu_\mathrm{L}$ depends on the magnetization direction $\vect{m}$. In contrast, the spin moments can be calculated neglecting spin-orbit coupling, since they are determined by the exchange interaction, which is much larger than SOC. Thus, including SOC in the calculation of $\mu_\mathrm{S}$, leaves the results practically unchanged. From simple arguments, the ratio between the orbital and spin moments can be estimated to be $\mu_\mathrm{L}/\mu_\mathrm{S} \propto \xi \propto Z^2$, with the spin-orbit strength $\xi$ and atomic number $Z$. Taking the values of Table \ref{TableMagMom}, indeed, the ratios for the two atom types yield $(\mu^{@\mathrm{W1}}_\mathrm{L}/\mu^{@\mathrm{W1}}_\mathrm{S})/(\mu^{@\mathrm{Cr}}_\mathrm{L}/\mu^{@\mathrm{Cr}}_\mathrm{S}) \approx 8 \pm 5$, which yields the same order of magnitude as the simple estimation by $Z^2_\mathrm{W}/Z^2_\mathrm{Cr} = 9.5$.


\subsection{Spin Stiffness}
\label{chap:results.spinstiffness}
In order to extract the spin-stiffness constant $A$ by a quadratic fit to the dispersion relation ({\it cf.}\ Eq.~\eqref{dispersion}), we calculate homogeneous spin spirals with various period lengths $\lambda$ employing the force theorem and neglecting SOC. In this way, we only have to perform self-consistent calculations for the antiferromagnetic state ($q'=0$).

\begin{figure}[htb]
  \includegraphics[width=85mm]{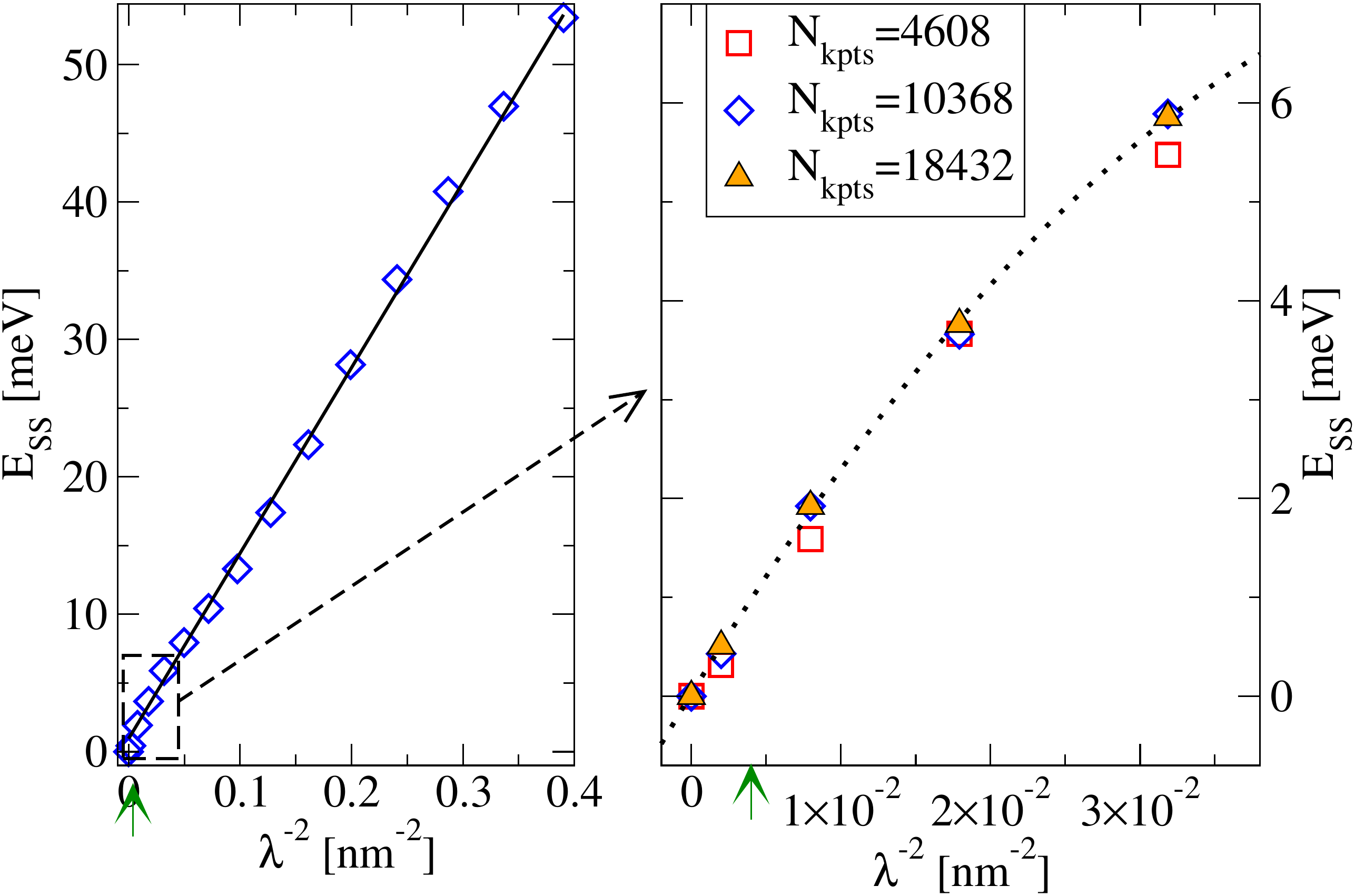}
  \caption{(Color online) Energy dispersion $E_\mathrm{SS}$ for spin spirals along the $[001]$ direction. Note that $\lambda^{-2}$ is used as scale of the abscissa. The arrow indicates the pitch corresponding experiment. Left panel: Calculated energies over a large interval of $q \sim \lambda^{-1}$ with 10368 $\vect{k}$-points in the full zone. The linear behavior in $\lambda^{-2}$ is modulated by a small oscillation. Right panel: Zoomed region. The shape of the curve is independent on the number of $\vect{k}$-points.}
  \label{fig:stiffness}
\end{figure}

The dispersion relations for spin spirals along the $[001]$ direction for different $\vect{k}$-point sets are shown in Fig.~\ref{fig:stiffness}. The left panel shows the results obtained for 10368 $\vect{k}$-points in the full two-dimensional Brillouin zone (symbols) and a linear fit in $\lambda^{-2}$ resulting in $A_{[001]} = 135\,\mathrm{meV}\,\mathrm{nm^2}$ (solid line). Relative to this linear behavior of the energy in $\lambda^{-2}$, a small oscillatory deviation is observed. Zooming into the region of interest as indicated by the arrow representing the experimentally observed modulation of magnetic contrast, reveals a curved function, whose shape is independent on the number of $\vect{k}$-points. Fitting a linear curve in $\lambda^{-2}$ to the small $\vect{q}$ data (first three points in the right panel of Fig.~\ref{fig:stiffness}) results in an enhanced spin-stiffness of  $A_{[001]}' \approx 235 \, \mathrm{meV\,nm^2}$. For spin spirals along $[1\overline{1}0]$ similar features were obtained (not shown) with $A_{[1\overline{1}0]} = 112\,\mathrm{meV}\,\mathrm{nm^2}$ and $A_{[1\overline{1}0]}' \approx 75 \, \mathrm{meV \, nm^2}$. However, we assume that these oscillations are unphysical, and they are effectively averaged out by choosing as spin stiffness constant $A_{[001]}$ and $A_{[1\overline{1}0]}$ for further considerations.

\subsection{DM-Vector}

\begin{figure}[htb]
  \includegraphics[width=85mm]{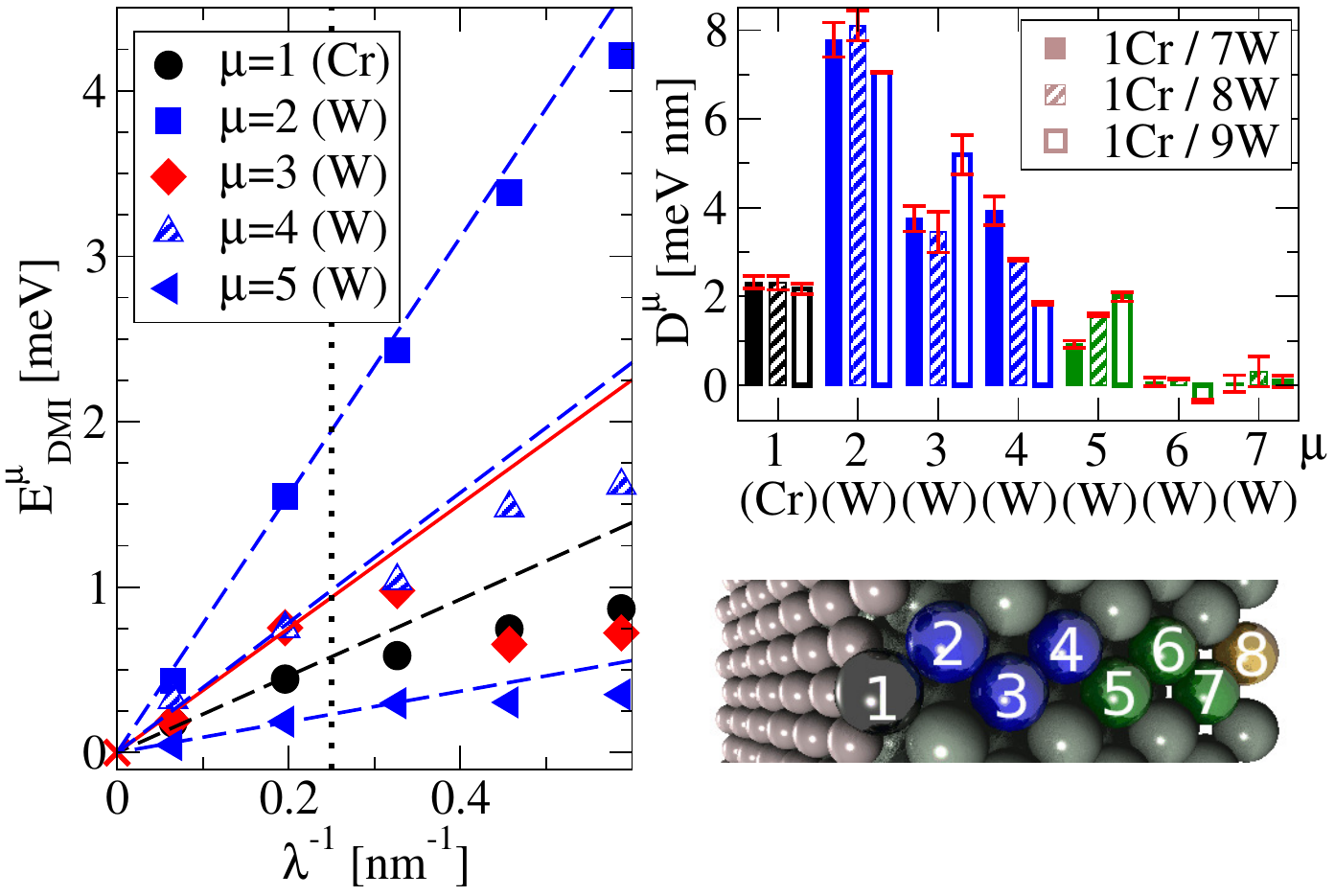}
  \caption{(Color online) Left panel: Layer-resolved DMI energy for spin spirals along the $[001]$ direction for various period lengths and linear fits to obtain the parameter $D^{\mu}$. Upper right: The distribution of the DMI among the layers of films consisting of 7, 8, and 9 W-layers. The layers are labeled as shown in the lower right panel for a film with 7 W-layers. The error bars in the upper right panel represent uncertainties of the fits.}
  \label{fig:ResultsDMI}
\end{figure}

To determine the Dzyaloshinskii-Moriya interaction, we calculate the energy of spin spirals with spin-orbit interaction for various period lengths. We first calculate in the p(1$\times$1) unit cell the magnetization- and charge density for the antiferromagnetic state in scalar-relativistic approximation (SRA) self-consistently. Next, for a spin spiral with various wave vectors $\vect{q}$ we find the Kohn-Sham eigenstates in SRA, constructing the DFT Hamiltonian from the same AFM charge- and magnetization density (force theorem). Finally, for each $\vect{q}$ we calculate the changes due to SOC on the Kohn-Sham eigenvalues in first-order perturbation theory. Following the procedure described in Sec.~\ref{sec:meth:DMI}, we obtain a layer-resolved analysis of the DMI energies, as shown in the left panel of Fig.~\ref{fig:ResultsDMI} for spirals along the $[001]$ direction. We extract the layer-resolved parameters $D^\mu$ through a linear fit ($E_\mathrm{DMI}^{\mu} = D^{\mu} \, \lambda^{-1}$) to the energies for long-period lengths ($\lambda \geq 4\,\mathrm{nm}$, {\it i.e.}\ the points to the left of the dotted line in the left panel of Fig.~\ref{fig:ResultsDMI}) and show in the upper right panel the resulting distribution of the layer-resolved $D^{\mu}$ to the total DMI strength $D = \sum_\mu D^{\mu}$ summarized in Table~\ref{totalDMI}. For comparison, films with seven to nine W layers are shown.

The total DMI-strength $D$ and all its layer-resolved contributions $D^\mu$ are positive. That means for Cr/W(110) the DMI favors a left-rotating spiral.
The first three W layers contribute about 80\% to the total DMI strength, the W interface layer yielding the major contribution. This is explained by the induced magnetic spin- and orbital moments due to hybridization with the Cr layer and the large spin-orbit coupling strength $\xi$ of W, which enter Eq.~\eqref{SOCoperator}. However, following this argument, the spin- and orbital moments do not enter linearly in $D^{\mu}$, as can be concluded by comparing the value of the magnetic moments for W ({\it cf.}\ Tab.~\ref{TableMagMom}) with the distribution in Fig.~\ref{fig:ResultsDMI}: the magnetic moments decay much faster with increasing distance from the Cr covered surface than the DMI. In Ref.~\onlinecite{zigzag} a minimal model is developed for the DMI which shows that the DMI depends on the hybridization of spin-orbit active orbitals on the W site with spin-dependent orbitals on the Cr site, and this is a more non-local effect. 

Spin spirals along the $[1\overline{1}0]$ direction have a smaller DMI strength $D_{[1\overline{1}0]}$ ({\it c.f.}\ Table~\ref{totalDMI}) and the contributions decay slower from the surface with increasing distance. Interestingly, in this case the Cr layer does not contribute significantly to the DMI (not shown).

Comparing the layer distribution for films of different thickness, it is observed that the distribution changes by choosing a different number of substrate layers, but the sum over all significant contributions ({\it i.e.}\ the first 5 layers for $[001]$ and 7 layers for $[1\overline{1}0]$) is nearly constant ({\it cf.}\ Tab.~\ref{totalDMI}).

\begin{table}
 \begin{ruledtabular}
  \begin{tabular}{r|cc}
   $[\mathrm{meV}\,\mathrm{nm}]$ & $[001]$ &  $[1\overline{1}0]$ \\[1.2ex] \hline
   1Cr / 7W & 19.9 & 8.5 \\
   1Cr / 8W & 19.3 & 8.5 \\
   1Cr / 9W & 19.0 & 8.8
  \end{tabular}
 \end{ruledtabular}
 \caption{Total DMI strength $D$ for spin spirals along a high-symmetry direction for films of different thicknesses.}
 \label{totalDMI}
\end{table}

\begin{figure}[htb]
  \includegraphics[width=75mm]{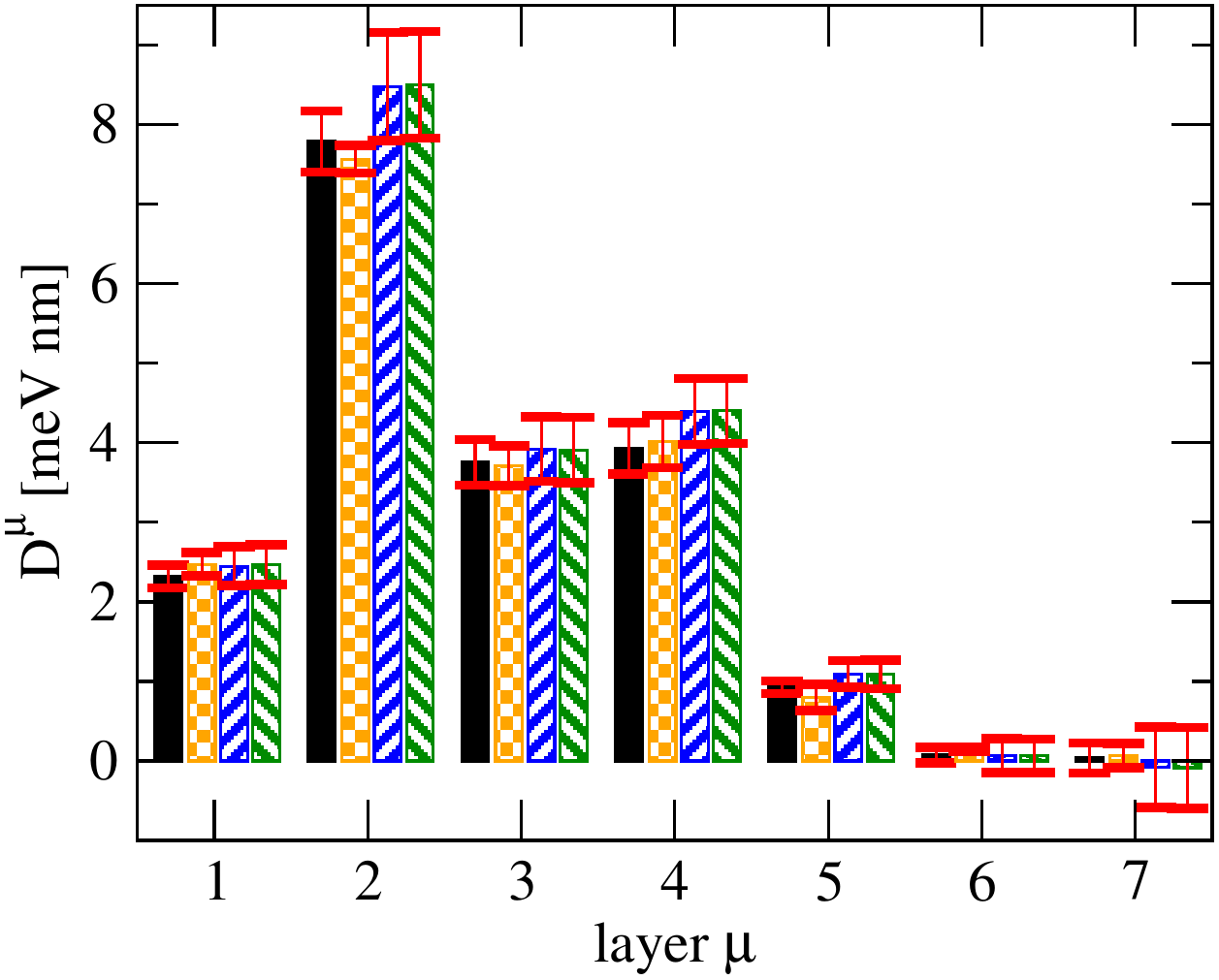}
  \caption{(Color online) Effects of different approximations for the initial Kohn-Sham states applied in Eq.~\eqref{FstO_expctval} on the layer-resolved contribution to DMI: Excluding SOC in the self-consistent calculation and employing the magnetic force theorem (FT) for the spin spirals (black, filled), excluding SOC but carrying out self-consistent calculations for the spin spirals  (orange, checkerboard), and including SOC with $\vect{m}_0$ along the $[001]$ and $[110]$ directions and employing the FT (blue, striped diagonally topright and green, topleft, respectively).}
  \label{fig:ApproxDMI}
\end{figure}

Our approach consists of two basic assumptions, namely that (a) the force theorem can be applied for the calculations of the spin-spiral energy starting from the charge density of the anti-ferromagnetic state in the p(1$\times$1) unit cell ignoring the spin-orbit interaction and (b) the results depend only weakly on the approximation that SOC is neglected in the calculation of the AFM initial charge density. To investigate the former assumption, we calculated a self-consistent charge density (excluding SOC) for each $\vect{q}$-vector and treated only SOC as perturbation. The changes with respect to the reference calculation ({\it cf.}\ full and checkerboard bars in Fig.~\ref{fig:ApproxDMI}) show that the magnetic force theorem is a very reasonable approach. To test the latter assumption, we included SOC in the self-consistent calculation of the AFM initial charge and magnetization density with a magnetization density $\vect{m}_0$ pointing along one of the high-symmetry orientations $\uvect{e}_k$, and calculate the DMI via the force theorem following the procedure as explained at the beginning of this section. Note, these calculations require the explicit use of the antiferromagnetic c(2$\times$2) unit cell containing 2 surface Cr atoms. The layer-resolved contribution to the DMI gets slightly enhanced in the W layers, but the result does not depend on the direction of the initial magnetization $\vect{m}_0$. 

\section{Discussion}

\begin{table*}
 \begin{ruledtabular}
  \begin{tabular}{rcc c d{-1}d{-1}c d{-1}ccd{-1}}
  3$d$& short-range & $\uvect{e}_\mathrm{easy}$ & $\uvect{q}$                & \multicolumn{1}{c}{$A$}   & \multicolumn{1}{c}{$D$} & $K$ & \multicolumn{1}{c}{$\kappa$} & \multicolumn{1}{c}{$\lambda$} & \multicolumn{1}{c}{$\lambda_\mathrm{hs}$} & \multicolumn{1}{c}{$E_\mathrm{hs}$} \\
      & order &                    &                    & \multicolumn{1}{c}{$[\rm meV \, nm^2]$} & \multicolumn{1}{c}{$[\rm meV \, nm]$} &  \multicolumn{1}{c}{[meV]} & & \multicolumn{1}{c}{[nm]} & \multicolumn{1}{c}{[nm]} & \multicolumn{1}{c}{[meV]} \\[1.3ex] \hline
  Cr  & AFM   & $[110]$            & $[001]$            & 135  & 19.9 & $0.9 \pm 0.1$ & 0.5                      & $-14.3$  & $-13.5$  & -0.3  \\
      &       &                    & $[1\overline{1}0]$ & 112  &  8.5 & $1.2 \pm 0.1$ & 3.0                      & $\infty$ & $\infty$ &  0.4  \\[1.7ex]
  Mn  & AFM   & $[1\overline{1}0]$ & $[1\overline{1}0]$ & 94.2 & 23.8 & 1.2           & 0.3                      & $-8.0$   & $-7.9$   & -0.9  \\[1.7ex]
 Fe-ML& FM    & $[1\overline{1}0]$ & $[001]$            & 51   & -2.6 & $^*$          & \multicolumn{1}{c}{$^*$} & $\infty$ & $\infty$ &  2.5  \\
      &       &                    & $[1\overline{1}0]$ & 131  &  7.4 & 2.5           & 9.7                      & $\infty$ & $\infty$ &  1.0  \\[1.7ex]
 Fe-DL& FM    & $[110]$            & $[001]$            & 165  & -3.6 & 0.1           & 2.1                      & $\infty$ & $\infty$ &  0.03 \\
      &       &                    & $[1\overline{1}0]$ & 143  &  3.1 & 0.2           & 4.8                      & $\infty$ & $\infty$ &  0.08 \\[1.2ex]
  \end{tabular}
 \end{ruledtabular}
 \caption{Model parameters $A$, $D$ and $K$ (spin stiffness, DMI and MAE, respectively) as well as the direction of the easy axis, $\uvect{e}_\mathrm{easy}$, as extracted from \textit{ab initio} calculations, resulting inhomogeneity parameter $\kappa$ and period length $\lambda$ of spin spirals along a chosen direction of the wave vector $\uvect{q}$ for different $3d$ ultra-thin films deposited on a W(110) substrate. The model parameters for a Mn monolayer and Fe doublelayer (Fe-DL) are taken from Refs.~\onlinecite{Bode:07.1} and \onlinecite{Heide:08.1}, respectively, and the parameters (except $D$, see text) for Fe monolayer (Fe-ML) are taken from Ref.~\onlinecite{Heide:06.1}. For films with p(1$\times$1) FM short-range order, the values for $K$ include also the contribution from the dipole-dipole interaction. For comparison also the period length $\lambda_\mathrm{hs}$ and energy $E_\mathrm{hs}$ (with respect to the collinear state) are given, where the model is restricted to homogeneous spin spirals. A negative sign of the period length indicates a left-rotational sense. The asterisk denotes that the easy axis is parallel to the Dzyaloshinskii-Moriya vector, and the micromagnetic model \eqref{eq:micmod} needs to be extended as explained in Ref.~\onlinecite{Heide:11.1}}.
 \label{finaltablecrw}
\end{table*}

Table~\ref{finaltablecrw} summarizes the three parameters $A$, $D$, and $K$ for the system Cr/W(110) as extracted from the \textit{ab initio} calculations. Employing the micromagnetic model for inhomogeneous spin spirals, Eq.~\eqref{micmod:flatspiral}, we find that the dimensionless inhomogeneity parameter $\kappa=16/\pi^2\, AK/D^2$ defined in \eqref{kappa:definition}, which is essential in the criterion for the energetic stability of a spin-spiral state, $\kappa\in [0,1[$, is smaller than one ($\kappa = 0.5 < 1$) and a left-rotating cycloidal spin spiral along the $[001]$ direction with a period length of 14.3~nm (corresponding to 45 chemical p(1$\times$1) unit cells in this direction) is energetically stable and is the magnetic ground state. In contrast, for spin spirals along the $[1\overline{1}0]$ direction, $\kappa$ is larger than one ($\kappa = 3.0 > 1$) and a periodic spiral is not stable. This is due to a large anisotropy of $D$, being about twice as large for spin spirals along $[001]$ compared to the ones along the $[1\overline{1}0]$ direction, whereas $A$ and $K$ only depend weakly on the propagation direction $\uvect{q}$. Moreover, $D$ enters quadratically in the criterion for the formation of a spin-spiral ground state. This quadratic dependence also appears in the energy gain over the AFM state for homogeneous spirals, and the strong $D$ for $\vect{q} \parallel [001]$ pushes this energy gain to about 0.3~meV per surface atom over the AFM state. In contrast, a spin spiral along the $[1\overline{1}0]$ direction is energetically unfavorable by 0.4~meV over the AFM.

The inhomogeneity of the ground-state spin spiral along the $[001]$ direction is considerable, as can be seen by comparing the period lengths. It is about $6\%$ larger than the corresponding homogeneous spiral.

\begin{figure}[htb]
  \includegraphics[width=85mm]{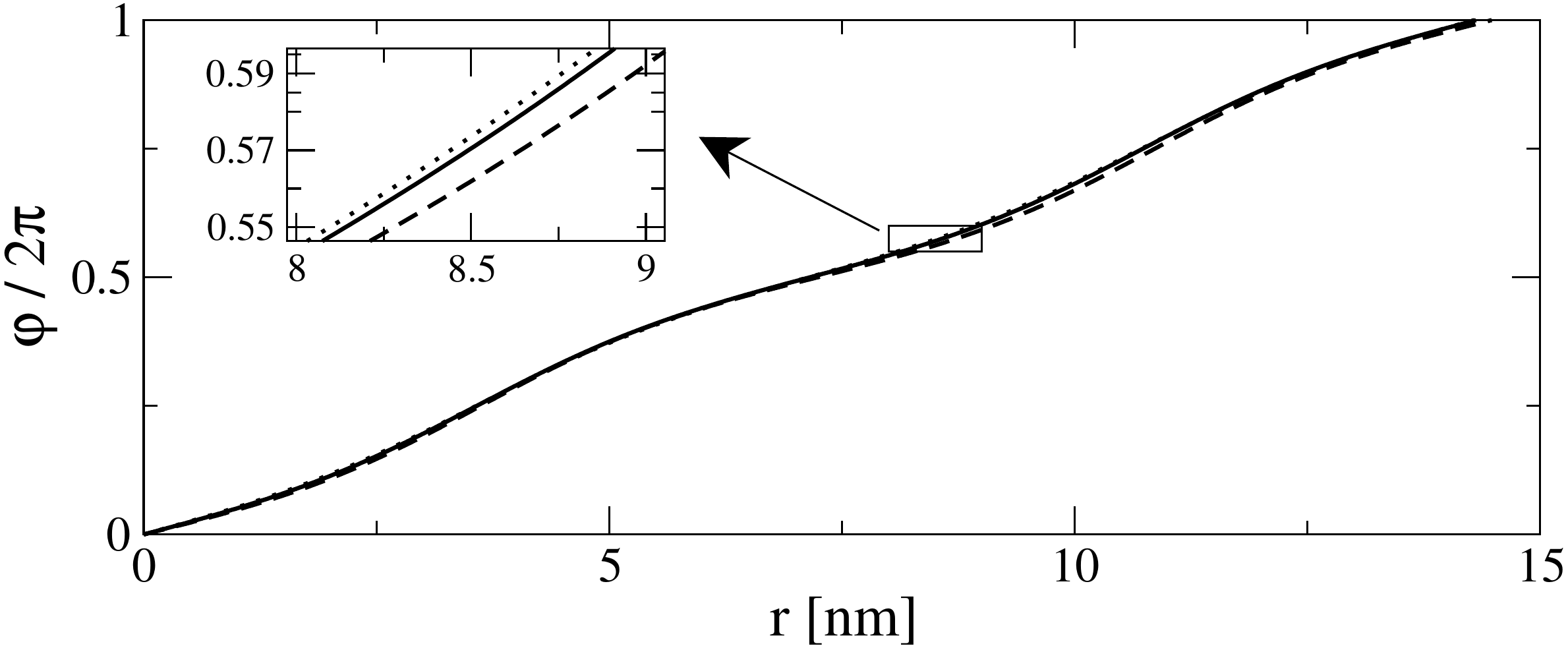}
  \caption{Profile of the predicted inhomogeneous spin spiral along $[001]$ with $A=135 \, \mathrm{meV\,nm^2}$, $D=19.9\, \mathrm{meV\, nm}$ and $K = (0.8, \, 0.9, ~ \mathrm{and} \, 1.0)\,\mathrm{meV}$, resulting in $\kappa=0.45$ (dotted line), $0.50$ (solid) and $0.55$ (dashed). The period length and the profile are not very sensitive to the uncertainty in $K$.}
   \label{fig:spiralprofile}
\end{figure}

The uncertainty in $K$ has only a minor influence on the profile and the period length of the spiral, as can be seen in Fig.~\ref{fig:spiralprofile}. There is a significant influence of the method chosen to obtain the spin stiffness, $A$, as discussed in Sec.~\ref{chap:results.spinstiffness}. Would we use the unphysically larger value $A'$ in the micromagnetic model, $\kappa$ increases to a value of $0.9$, accompanied by a strong increase of the spiral period ({\it cf.}\ Fig.~\ref{fig:inhom}).

\begin{figure}[htb]
  \includegraphics[width=85mm]{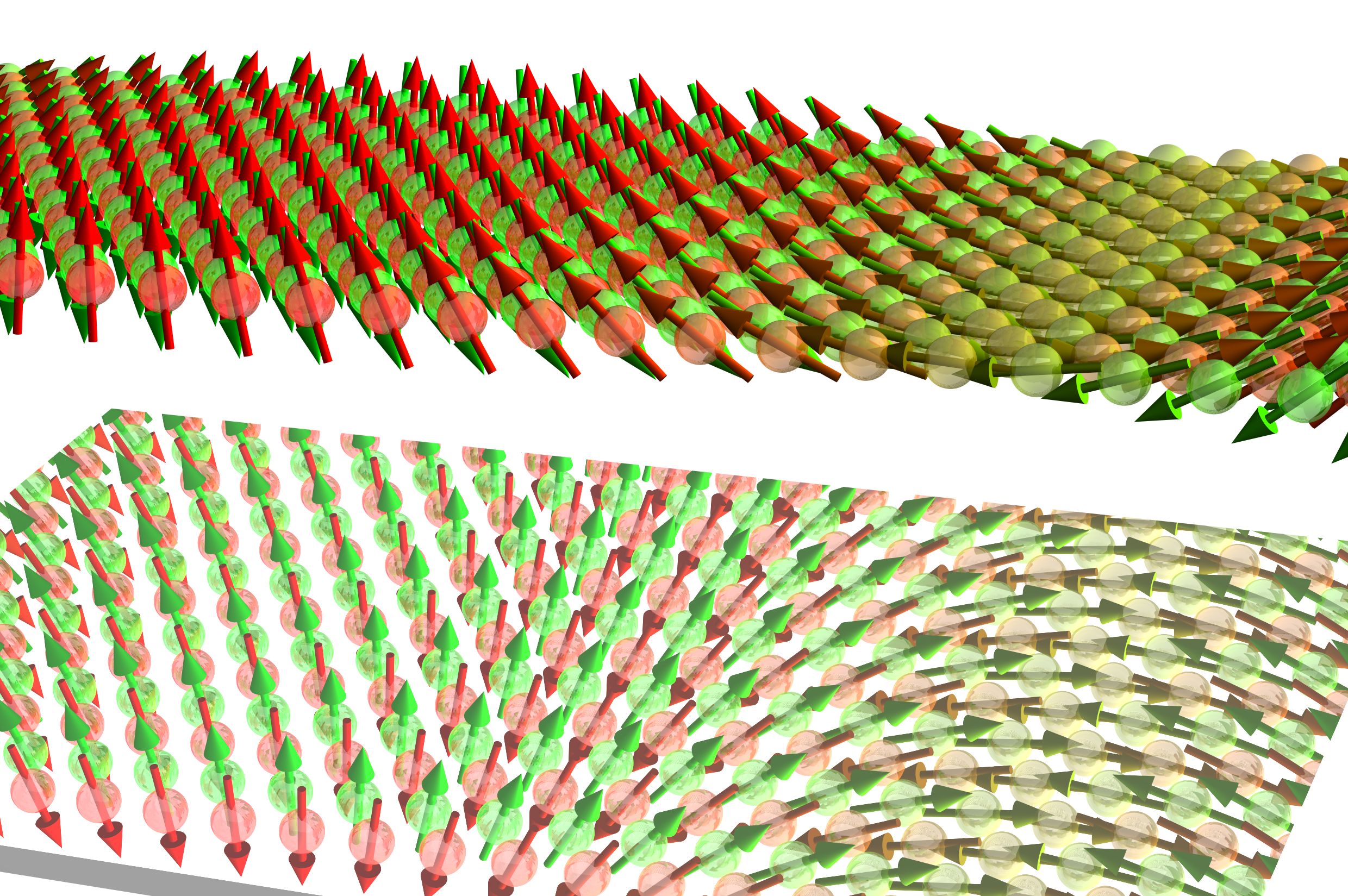}
  \caption{(Color online) Upper figure: Predicted left rotating magnetic spin-spiral ground state of the Cr monolayer on W(110) shown for roughly $1/4$th of the period length. The arrows represent the magnetization direction of the Cr atoms (the W-substrate atoms are not shown, but assumed to be below the Cr-layer). The color represents the out-of-plane component of the magnetization direction, ranging from red ($m_z=+1$) via brown ($m_z=0$) to green ($m_z=-1$). Lower figure: The mirror image shows a right rotating spiral that is energetically unfavorable and not found in nature.}
   \label{fig:spiral_sketch}
\end{figure}

\subsection{Comparison to the Experiment}

Our results can explain the modulation of magnetic contrast along the $[001]$ direction as found in the SP-STM experiment by Santos \etal\cite{Santos:08.1} in terms of spin spirals, which are driven by the DMI: (i) We have an excellent agreement between the experimentally observed propagation direction of the spin spiral, and (ii) the modulation length of $7.7 \pm 0.5~\mathrm{nm}$ (compared to our findings of $|\lambda| / 2 = 7.2~\mathrm{nm}$), giving confidence in our theoretically determined parameter set $A$, $D$, $K$. A sketch of the spiral is shown in Fig.~\ref{fig:spiral_sketch}.


\subsection{Comparison to other Thin-Film Systems on W(110) Substrates}

In order to get  a better understanding on the intricate behavior of the DMI in competition to the spin-stiffness and MAE through a systematic analysis we compare our findings to other $3d$ transition-metal thin-film systems on W(110) substrates ({\it cf.}\ Table~\ref{finaltablecrw}). We find that the Cr/W(110) system is closest to Mn/W(110).\cite{Bode:07.1} Both systems exhibit a short-range antiferromagnetic order and produce a high DMI with the same sign that creates a spin-spiral ground state of the same rotational sense and comparable pitch. However, the propagation direction is different, along $[1\overline{1}0]$ for Mn/W(110) and along $[001]$ for Cr/W(110). Although both systems  are so similar, the question arises, why are the propagation directions of the spin-spiral ground states different? It turns out that the direction of the easy axis has a crucial influence on the energy of the spin spiral: For Mn/W(110), the easy axis lies in-plane along the $[1\overline{1}0]$-direction. A possible flat spin spiral along the $[001]$ direction would rotate in a plane spanned by the $[001]$ and $[110]$-directions, and thus would \emph{exclude} the easy axis. Therefore, a large anisotropy energy needs to be overcome to rotate the magnetic moments away from the easy axis into the rotational plane. For the case of Mn/W(110), these costs cannot be fully compensated by a gain in energy through the DMI. Thus, in Mn/W(110) the direction of the easy axis prohibits a spiral-formation in $[001]$-direction. In contrast, in Cr/W(110) the energy gain due to DMI is of similar magnitude as compared to Mn/W(110), but the easy axis points out-of-plane, and thus the rotational plane of a spin spiral \emph{includes} the easy axis, irrespective of the propagation direction of the spin spiral. As explained above, for Cr/W(110) the DMI strength for spin spirals in the $[001]$-direction is about twice as large as in the $[1\overline{1}0]$-direction, and thus determines the propagation direction of the spiral.

According to the theoretical values presented in Table~\ref{finaltablecrw}, in the Fe mono- (Fe-ML) and doublelayer (Fe-DL) on W(110), the DMI is too weak to compete against the MAE and spin stiffness, and the ground states of both remain ferromagnetic, but the DMI influences the dynamical properties in terms of the excitation spectra of magnons.\cite{Udvardi:2009, Zakeri:2010} In addition, for the Fe-DL Heide \textit{et al.}\cite{Heide:08.1} could explain by the micromagnetic model that the DMI is strong enough to change the domain wall structure from the Bloch wall to a N\'eel-type domain wall, to determine the orientation of the wall and the right rotating sense of the magnetization in the wall.
These walls are normal to the $[001]$ direction, {\it i.e.}\ the magnetization has the same propagation direction as for Cr/W(110), but the rotational sense is opposite. However, spin spirals can occur in stripes of Fe doublelayers of finite width.\cite{Meckler:2012}
\begin{figure}[htb]
  \includegraphics[width=85mm]{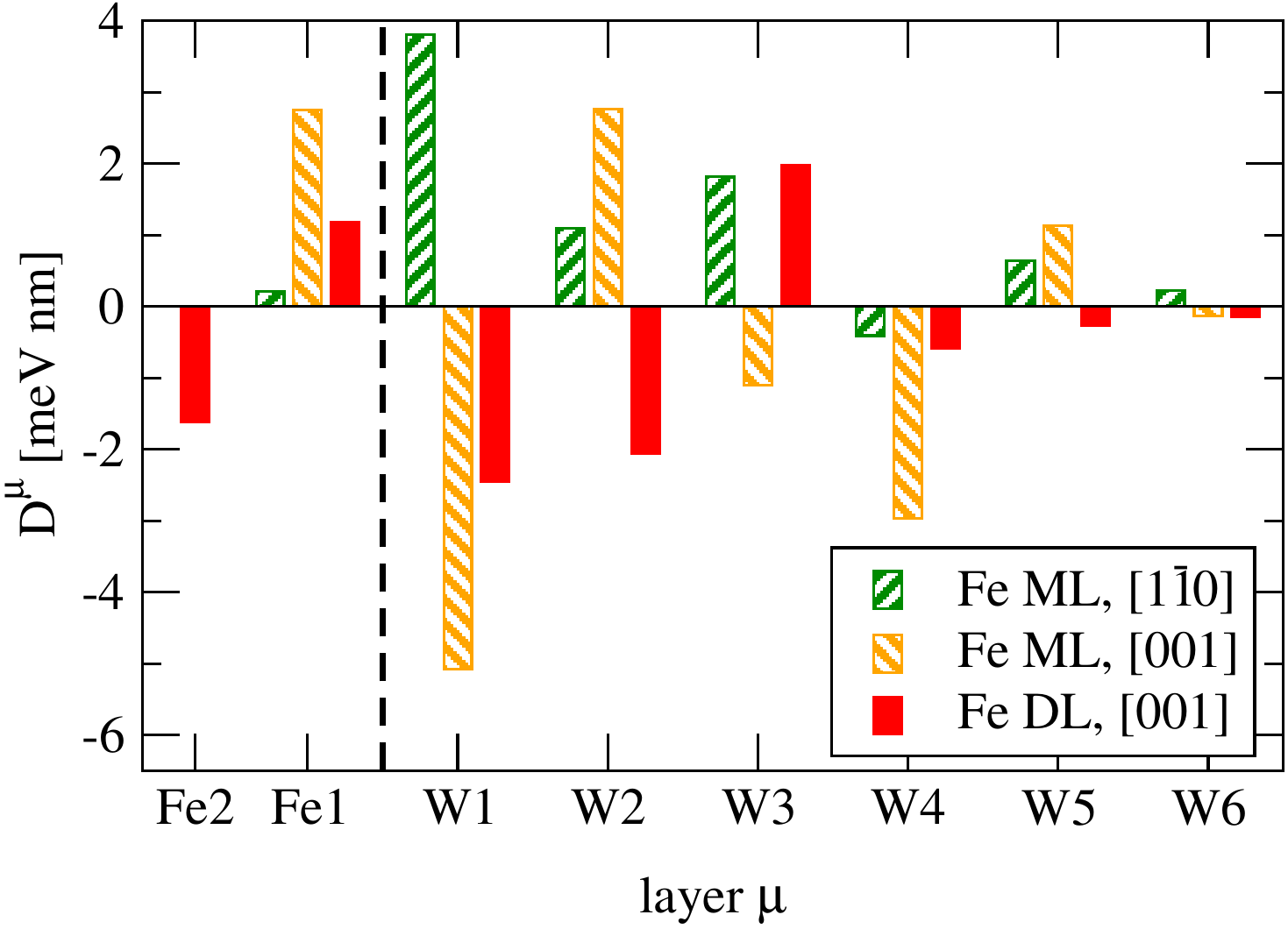}
  \caption{(Color online) Layer-resolved DMI strength for spin spirals in ferromagnetic Fe mono- (ML) and doublelayers (DL) on W(110).}
   \label{fig:DMI_FeW110}
  \end{figure}

In addition, we calculated the layer-resolved DMI for an Fe-ML and Fe-DL on W(110) with the computational parameters taken from Ref.~\onlinecite{Heide:09.1}, but using a slab of 7 W layers and 3 (7) $\vect{q}$-vectors with $\lambda^{-1} < 0.03\,a^{-1}$ ($\lambda^{-1} < 0.1\,a^{-1}$) for the Fe-ML (Fe-DL). The full Brillouin zone was sampled by 3600 and 3780 $k$-points for spin spirals along the $[1\overline{1}0]$ and $[001]$-direction, respectively. In the light of the Cr/W(110) results, where all layer-resolved DMI strengths $D^{\mu}$ had the same sign irrespective of the spiral direction, the resulting histograms (Fig.~\ref{fig:DMI_FeW110}) reveal a rather diverse behavior of $D^{\mu}$ as function of the layer index. Common to all systems studied here, \textit{i.e.}\ the Fe-ML, Fe-DL, Mn on W(110) (not shown here) and Cr on W(110) is the observation that the W interface layer (W1) shows the largest contribution to $D$. This can simply be attributed to the fact that the spin-polarized states of the magnetic thin film hybridize and thus spin-polarize W1 most, which experiences in addition the strongest break of the structure-inversion symmetry and a large SOC strength. However, the sign might crucially depend on both, the type of the magnetic thin film, and the direction of the spin spirals. For example, in case of the Fe-ML system, we witness a large negative contribution to $D$ at the W interface layer (W1) for spin spirals along the $[001]$-direction, while the W1 layer exhibits a large positive contribution to $D$ for spin spirals in the $[1\overline{1}0]$-direction. For spirals along the $[001]$ direction, the sign of $D^{\mu}$ for Fe and W atoms at the Fe-W interface is the same, irrespective whether one deals with an Fe-ML or Fe-DL. The values of $|D^\mu|$ of the interface atoms Fe1 and W1 (see Fig.~\ref{fig:DMI_FeW110}) are smaller in the Fe-DL presumably due to a smaller structural asymmetry. Most $D^{\mu}$ in the Fe-ML system for spin spirals along the $[1\overline{1}0]$-direction are of the same positive sign. This situation is closest to Cr/W(110), where all layers contribute with the same positive sign to $D$ (see Fig.~\ref{fig:ResultsDMI}), however irrespective whether the frozen spiral propagates along $[001]$ or $[1\overline{1}0]$ direction. This is different for the Fe-ML and Fe-DL  where in both cases the  $D^{\mu}$  show an oscillatory behavior as function of the layer number from the interface.  This shows again that the DMI is a non-local effect, and the DMI in the substrate atoms is drastically influenced by the surface layer.

\section{Summary}
We analyzed the magnetic structure of a monolayer Cr/W(110) by means of density functional theory. We proposed a minimal atomistic spin-model that describes the magnetic structure of the system. Following symmetry arguments we developed out of the spin-model a simple one-dimensional micromagnetic model and found the model parameters by fits to \textit{ab initio} calculations. Within the model, we are able to explain the experimentally observed modulation of magnetic contrast in terms of cycloidal spin spirals along the $[001]$ direction. We predict a chiral magnetic structure, which is an inhomogeneous spin spiral of left-handedness caused by the spin-orbit driven Dzyaloshinskii-Moriya interaction (DMI). The propagation direction of the frozen spiral and the period length of 14.3~nm agree nicely with the experimental values. Analyzing the layer-resolved distribution of the DMI, we can address the major contribution to the first W interface-layer originating from the large spin-orbit coupling strength, the induced magnetic moment and the structure-inversion asymmetry of the interface. However, it would be too simplistic to conclude that the interface atoms of a substrate with large spin-orbit interaction and thus the substrate as a whole determines chiral properties of the magnetic film. Instead the layer-distribution of the Dzyaloshinskii-Moriya strength (\textit{e.g.}\ the sign) differs qualitatively when comparing different thin-film systems on W(110), resulting in {\it e.g.}\ different strengths of the total DMI for different propagation directions. This reveals the complex interplay between chemical and structural factors responsible for the Dzyaloshinskii-Moriya interaction. 

Future systematic DFT studies of relativistic effects in thin film systems should help gaining a deeper understanding of the direction and the rotational sense of the spin-spiral ground state on the basis of the electronic structure of these interface stabilized two-dimensional chiral magnets. We invite experimental studies for Cr/W(110) to verify the handedness and the inhomogeneity in the rotation of the spin-spiral state.

\begin{acknowledgments}
We thank Kirsten von Bergmann for making us aware of the experimental results by Santos \textit{et al.}, and Benedikt Schweflinghaus and Paolo Ferriani for fruitful discussions. We gratefully acknowledge funding under HGF YIG Program VH-NG-513, as well as computing time on the supercomputers JUQUEEN and JUROPA at J\"{u}lich Supercomputing Centre and JARA-HPC of RWTH Aachen University.
\end{acknowledgments}

\appendix \section{Matrix Elements of Spin-Orbit Operator for Spin-Spiral States} \label{sec:appendix}

According to the generalized Bloch theorem, the Bloch state
\begin{equation}
  \psi_{\vect{k}\nu} ({\vect r}|{\vect q}) = \left( \begin{array}{c} \expo{i (\vect{k}-\vect{q}/2) \cdot \vect{r}} ~ u^\uparrow_{\vect{k}\nu} ({\vect r}) \\ \expo{i (\vect{k}+\vect{q}/2) \cdot \vect{r}} ~ u^\downarrow_{\vect{k}\nu} ({\vect r}) \end{array} \right) \,,  
\end{equation}
is an eigenstate of the scalar-relativistic Hamiltonian for a homogeneously spiraling magnetic structure ({\it cf.}\ Eq.~\eqref{GenBlochTheorem})
with functions $u^\uparrow_{\vect{k}\nu}$ and $u^\downarrow_{\vect{k}\nu}$ having the periodicity of the chemical lattice, {\it i.e.}\ the lattice when magnetism is ignored. All four components of the spin-orbit operator also exhibit the periodicity of the chemical lattice, which implies that the action of those on a Bloch state returns the same exponential factor, but a different lattice periodic function $\tilde{u}$,
\begin{equation}
  \matr{H}_\mathrm{so}^{\sigma' \sigma} ~ \expo{i (\vect{k} \mp \vect{q}/2) \cdot \vect{r}} ~ u^\sigma_{\vect{k}\nu} ({\vect r}) =  \expo{i (\vect{k} \mp \vect{q}/2) \cdot \vect{r}} ~ \tilde{u}^{\sigma' \sigma}_{\vect{k}\nu} ({\vect r})~.
\end{equation}
In the following, $\vect{R}$ denotes the shortest possible lattice vector in the direction of $\vect{q}$ and defines a unit-cell volume $\Omega_\textrm{R}$. We further choose the modulus of $\vect{q}$ such that $\vect{q} \cdot \vect{R} = 2\pi/N$, with integer $N$ denoting the length of the magnetic supercell in units of $\abs{\vect{R}}$. The expectation value consists of four contributions, $\langle \psi_{\vect{k}\nu} \lvert \mathcal{H}_\mathrm{so} \rvert \psi_{\vect{k}\nu} \rangle = \sum_{\sigma, \sigma'} \langle \psi^{\sigma}_{\vect{k}\nu} \lvert \mathcal{H}^{\sigma\sigma'}_\mathrm{so} \rvert \psi^{\sigma'}_{\vect{k}\nu} \rangle$ ({\it cf.}\ Eq.~\eqref{FstO_expctval}). Integration over the magnetic supercell with volume $\Omega_N$ yields
\begin{eqnarray}
  \langle \psi^\uparrow_{\vect{k}\nu} \lvert \matr{H}_\mathrm{so}^{\uparrow\uparrow} \rvert \psi^\uparrow_{\vect{k}\nu} \rangle &=& \sum_{n=1}^{N} \int\limits_{\Omega_n} \mathrm{d}\vect{r} ~ u^{\uparrow^*}_{\vect{k}\nu} (\vect{r}+n\vect{R}) \, \tilde{u}^{\uparrow \uparrow}_{\vect{k}\nu} (\vect{r}+n\vect{R}) 
  \nonumber \\
  &=& N ~ \int\limits_{\Omega_\textrm{R}} \mathrm{d}\vect{r} ~ u^{\uparrow^*}_{\vect{k}\nu} (\vect{r}) \, \tilde{u}^{\uparrow\uparrow}_{\vect{k}\nu} (\vect{r}) ~, \label{app:deriv:upup}
\end{eqnarray}
(and a similar expression for $\langle \psi^\downarrow_{\vect{k}\nu} \lvert \matr{H}_\mathrm{so}^{\downarrow\downarrow} \rvert \psi^\downarrow_{\vect{k}\nu} \rangle$). The lattice periodicity of $u$ and $\tilde{u}$ has been employed from the first to the second line, yielding $N$ identical integrals which are restricted to the unit cell with volume $\Omega_\textrm{R}$. The matrix-elements of the spin-flip contribution of the spin-orbit operator vanish,
\begin{eqnarray}
  \langle \psi^\uparrow_{\vect{k}\nu} \lvert \matr{H}_\mathrm{so}^{\uparrow\downarrow} \rvert \psi^\downarrow_{\vect{k}\nu} \rangle &=& \sum_{n=1}^{N} \expo{i\,n \vect{q}\cdot \vect{R}} \nonumber \\ && \times \int\limits_{\Omega_n} \mathrm{d}\vect{r} ~ \expo{i\, \vect{q}\cdot \vect{r}} ~ u^{\uparrow^*}_{\vect{k}\nu} (\vect{r}+n\vect{R}) \, \tilde{u}^{\uparrow \downarrow}_{\vect{k}\nu} (\vect{r}+n\vect{R}) 
  \nonumber \\
  &=& \left[ \sum_{n=1}^{N} \expo{2\pi \, i \,n/N}\right] ~ \int\limits_{\Omega_\textrm{R}} \mathrm{d}\vect{r} ~ \expo{i\, \vect{q}\cdot \vect{r}} ~ u^{\uparrow^*}_{\vect{k}\nu} (\vect{r}) \, \tilde{u}^{\uparrow \downarrow}_{\vect{k}\nu} (\vect{r})  \nonumber \\
  &=& 0 ~, \label{app:deriv:updown}
\end{eqnarray}
(and similarly for $\langle \psi^\downarrow_{\vect{k}\nu} \lvert \matr{H}_\mathrm{so}^{\downarrow\uparrow} \rvert \psi^\uparrow_{\vect{k}\nu} \rangle$) due to the summation over the exponential factor, $\sum_{n=1}^N \mathrm{exp}(2\pi \, i\,n/N) = 0$. Thus, the total contribution can be evaluated in the unit cell restricted to $\Omega_\textrm{R}$. In our case of $\vect{q}$ along high-symmetry directions of the lattice, this $\Omega_\textrm{R}$ corresponds to the chemical p(1$\times$1) unit cell.

Please notice this holds only true for first-order perturbation theory. If one goes beyond first-order perturbation theory, also matrix elements of the form $\langle \psi^\uparrow_{\vect{k}+\vect{q}\nu'} \lvert \matr{H}_\mathrm{so}^{\uparrow\downarrow} \rvert \psi^\downarrow_{\vect{k}\nu} \rangle$ need to be considered.\cite{Heide:09.1}

\bibliography{bibzi}

\end{document}